%% file: main.tex
\documentclass[conference, 10pt]{IEEEtran}
\IEEEoverridecommandlockouts

\input{math}

\input{preamble}
\input{tools}

 \usepackage{fontawesome}

\tikzset{
    every node/.append style={font=\footnotesize},
    every label/.append style={font=\footnotesize}
}
\pgfplotsset{
    every axis legend/.append style={
        fill opacity=0.8,
        draw=none,
    },
    every axis/.append style={
        x grid style={darkgray!60},
        xmajorgrids,
        y grid style={darkgray!60},
        ymajorgrids,
    }
}

\newcommand{\esl}{\gls{er}\xspace}
\newcommand{\esls}{\glspl{er}\xspace}

\begin{document}

\title{How to Perform Distributed Precoding to Wirelessly Power Shelf Labels:\\ Signal Processing and Measurements
\thanks{The work is supported by the REINDEER project under grant agreement No.~101013425.\\ This paper is accepted at IEEE 25th International Workshop on Signal Processing Advances in
Wireless Communications (IEEE SPAWC 2024) and can be found in~\cite{Call2409:How}.}
}

\author{
    \IEEEauthorblockN{%
    Gilles Callebaut, Jarne Van Mulders, Bert Cox, Liesbet Van der Perre, Lieven De Strycker, François Rottenberg}%
    \IEEEauthorblockA{\textit{Department of Electrical Engineering, KU Leuven}, Belgium}
}

\maketitle

\defineauthors{TODO, gilles, francois, liesbet, lieven, bert, jarne}

\begin{abstract}
\Gls{wpt} has garnered increasing attention due to its potential to eliminate device-side batteries. With the advent of (distributed) \gls{mimo}, \gls{rf} \gls{wpt} has become feasible over extended distances. This study focuses on optimizing the energy delivery to \glspl{er} while minimizing system total transmit power. Rather than continuous power delivery, we optimize the precoding weights within specified time slots to meet the energy requirements of the \glspl{er}. Both unsynchronized (non-coherent) and synchronized (coherent) systems are evaluated. Our analysis indicates that augmenting the number of antennas and transitioning from an unsynchronized to a synchronized full phase-coherent system substantially enhances system performance. This optimization ensures precise energy delivery, reducing overshoots and overall energy consumption. Experimental validation was conducted using a testbed with 84 antennas, validating the trends observed in our numerical simulations. 

\end{abstract}


\input{sections/00_Introduction}
\input{sections/02_SignalProcessing}

\input{sections/03_Evaluation}

\input{sections/04_Measurements}

\input{sections/05_conclusions}

\printbibliography

\end{document}

%% file: math.tex
\usepackage{amsmath,amssymb,amsfonts}
\usepackage{algorithmic}

\usepackage{xifthen}
\newcommand{\prob}[1][]{%
\ifthenelse{\isempty{#1}}%
      {\ensuremath{P}}%
    {\ensuremath{P\left\(#1\right\)}}%
}

\newcommand{\vect}[1]{\boldsymbol{\mathrm{#1}}}
\newcommand{\mat}[1]{\boldsymbol{\mathrm{#1}}}

\newcommand{\tr}{\mathrm{tr}}

\usepackage{amsthm}
\theoremstyle{remark}

%% file: preamble.tex
\usepackage{graphicx} %
\graphicspath{{./figures}}
\usepackage[OT1]{fontenc}
\usepackage{color}
\usepackage[dvipsnames,svgnames, table]{xcolor} %
\usepackage[hidelinks]{hyperref} %
\usepackage{siunitx}
\DeclareSIUnit{\dBm}{dBm}	                %
\DeclareSIUnit{\dBi}{dBi}                   %
\DeclareSIUnit{\dBsm}{dBsm}                 %
\usepackage{booktabs}
\usepackage[backend=biber, style=ieee, citestyle=numeric-comp, isbn=false, doi=false,citecounter=true]{biblatex}

\usepackage{pgfplots}
\usepgfplotslibrary{groupplots,dateplot}
\usetikzlibrary{patterns,shapes.arrows}
\usetikzlibrary{spy}
\pgfplotsset{compat=newest}
\usetikzlibrary{external}
\usetikzlibrary{calc}
\usetikzlibrary{shadings}
\usetikzlibrary{shadows.blur}
\usetikzlibrary{decorations.pathreplacing}

\usepackage{soul}

\pgfmathdeclarefunction{viridis}{1}{%
  \pgfmathparse{%
    1000*(
    0.370489*pow(#1, 3)-
    1.11384*pow(#1, 2)+
    0.929777*(#1)
    )%
  }%
}

\usepackage{tabularx}
\usepackage{tablefootnote}

\AtBeginBibliography{\footnotesize}

\usepackage[font=small]{caption}
\usepackage{subcaption}
\usepackage[capitalize]{cleveref}
\usepackage{xspace}

\usepackage{placeins}
\usepackage[xindy, nonumberlist]{glossaries}
\makenoidxglossaries%
\loadglsentries{abbr}%

\usepackage{comment}

\newcolumntype{R}{>{\raggedleft\arraybackslash}X}

\usepackage{graphicx}
\usepackage{textcomp}
\usepackage{xcolor}
\def\BibTeX{{\rm B\kern-.05em{\sc i\kern-.025em b}\kern-.08em
    T\kern-.1667em\lower.7ex\hbox{E}\kern-.125emX}}

%% file: abbr.tex
\newacronym{2d}{2D}{two-dimensional}
\newacronym{3d}{3D}{three-dimensional}
\newacronym{3gpp}{3GPP}{3rd generation partnership project}
\newacronym{4g}{4G}{fourth-generation}
\newacronym{5g}{5G}{fifth-generation}
\newacronym{6dof}{6DoF}{six degrees of freedom}
\newacronym{6g}{6G}{sixth-generation}
\newacronym{abp}{ABP}{authentication by personalisation}
\newacronym{ac}{AC}{alternating current}
\newacronym{aclr}{ACLR}{adjacent channel leakage ratio}
\newacronym{adc}{ADC}{analog-to-digital converter}
\newacronym{adr}{ADR}{adaptive data rate}
\newacronym{afe}{AFE}{analog front end}
\newacronym{ag}{AG}{array gain}
\newacronym{ai}{AI}{artificial intelligence}
\newacronym{am}{AM}{amplitude modulation}
\newacronym{amam}{AM/AM}{amplitude modulation to amplitude modulation}
\newacronym{amc}{AMC}{automatic modulation classification}
\newacronym{amp}{AMP}{approximate message passing}
\newacronym{ampm}{AM/PM}{amplitude modulation to phase modulation}
\newacronym{aoa}{AOA}{angle-of-arrival}
\newacronym{aod}{AOD}{angle-of-departure}
\newacronym{ap}{AP}{access point}
\newacronym{api}{API}{application program interface}
\newacronym{apu}{APU}{access point unit}
\newacronym{ar}{AR}{augmented reality}
\newacronym{arp}{ARP}{Antenna Reference Point}
\newacronym{asic}{ASIC}{application specific integrated circuit}
\newacronym{ask}{ASK}{amplitude-shift keying}
\newacronym{auv}{AUV}{autonomous underwater vehicle}
\newacronym{awgn}{AWGN}{additive white Gaussian noise}
\newacronym{baw}{BAW}{bulk acoustic wave}
\newacronym{bb}{BB}{base-band}
\newacronym{bc}{BC}{backscatter communication}
\newacronym{bd}{BD}{backscatter device}
\newacronym{ber}{BER}{bit error rate}
\newacronym{bf}{BF}{beamforming}
\newacronym{bldc}{BLDC}{brushless DC}
\newacronym{bom}{BOM}{bill of materials}
\newacronym{bpsk}{BPSK}{binary phase-shift keying}
\newacronym{bs}{BS}{base station}
\newacronym{bu}{BU}{booster unit}
\newacronym{bw}{BW}{bandwidth}
\newacronym{ca}{CA}{carrier emitter}
\newacronym{cad}{CAD}{channel activity detection}
\newacronym{cars}{CARS}{calibration reference signal}
\newacronym{cbm}{CBM}{condition based maintenance}
\newacronym{cc}{CC}{constant current}
\newacronym{ccdf}{CCDF}{complementary cumulative distribution function}
\newacronym{ccnn}{CCNN}{circular convolutional neural network}
\newacronym{ccs}{CCS}{correlative channel sounder}
\newacronym{cdf}{CDF}{cumulative distribution function}
\newacronym{cdrx}{CDRX}{connected mode DRX}
\newacronym{ce}{CE}{coverage enhancement}
\newacronym{ced}{CED}{cumulative energy density}
\newacronym{cf}{CF}{cell-free}
\newacronym{cfmmimo}{CF-mMIMO}{cell-free massive MIMO}
\newacronym{cfo}{CFO}{carrier frequency offset}
\newacronym{cir}{CIR}{channel impulse response}
\newacronym{cla}{CLA}{closed-loop approach}
\newacronym{cmos}{CMOS}{complementary metal oxide semiconductor}
\newacronym{cost}{COST}{commercial off-the-shelf}
\newacronym{cots}{COTS}{commercial off-the-shelf}
\newacronym{cp}{CP}{cyclic prefix}
\newacronym{cpt}{CPT}{capacitive power transfer}
\newacronym{cpu}{CPU}{central-processing unit}
\newacronym{cqi}{CQI}{channel quality indicator}
\newacronym{cr}{CR}{coding rate}
\newacronym{crc}{CRC}{cyclic redundancy check}
\newacronym{crlb}{CRLB}{Cram\'er-Rao lower bound}
\newacronym{crs}{CRS}{cell reference signal}
\newacronym{cs}{CS}{compressed sensing}
\newacronym{cse}{CSE}{channel state estimation}
\newacronym{csi}{CSI}{channel state information}
\newacronym{csp}{CSP}{contact service point}
\newacronym{css}{CSS}{chirp spread spectrum}
\newacronym{cu}{CU}{central unit}
\newacronym{cv}{CV}{constant voltage}
\newacronym{cw}{CW}{continuous wave}
\newacronym{dab}{DAB}{Digital Audio Broadcasting}
\newacronym{dac}{DAC}{digital-to-analog converter}
\newacronym{daq}{DAQ}{data acquisition system}
\newacronym{das}{DAS}{distributed antenna systems}
\newacronym{dc}{DC}{direct current}
\newacronym{dcc}{DCC}{dynamic cooperation clustering}
\newacronym{de}{DE}{drain efficiency}
\newacronym{dft}{DFT}{discrete Fourier transform}
\newacronym{dl}{DL}{downlink}
\newacronym{dlc}{DLC}{Distributed Laser Charging}
\newacronym{dli}{DLI}{direct link interference}
\newacronym{dm}{DM}{diffuse multipath}
\newacronym{dma}{DMA}{Direct Memory Access}
\newacronym{dmc}{DMC}{diffuse multipath component}
\newacronym{dmimo}{D-MIMO}{distributed MIMO}
\newacronym{doa}{DOA}{direction-of-arrival}
\newacronym{dpd}{DPD}{digital pre-distortion}
\newacronym{dpdk}{DPDK}{Data Plane Development Kit}
\newacronym{dram}{DRAM}{dynamic random-access memory}
\newacronym{drx}{DRX}{Discontinuous Reception Mode}
\newacronym{dsl}{DSL}{digital subscriber line}
\newacronym{dsp}{DSP}{digital signal processing}
\newacronym{dss}{DSS}{Dataset Storage Standard}
\newacronym{duc}{DUC}{digital up-converter}
\newacronym{dvb}{DVB}{Digital Video Broadcasting}
\newacronym{e2e}{E2E}{end-to-end}
\newacronym{easa}{EASA}{European Union Aviation Safety Agency}
\newacronym{ec}{EC}{European Commission}
\newacronym{ecdf}{eCDF}{empirical cumulative distribution function}
\newacronym{ecsp}{ECSP}{edge computing service point}
\newacronym{edlc}{EDLC}{electrostatic double-layer capacitors}
\newacronym{edrx}{eDRX}{Extended Discontinuous Reception Mode}
\newacronym{ee}{EE}{energy efficiency}
\newacronym{egprs}{EGPRS}{Enhanced Data Rates for GSM Evolution}
\newacronym{eh}{EH}{energy harvesting}
\newacronym{eirp}{EIRP}{equivalent isotropically radiated power}
\newacronym{em}{EM}{electromagnetic}
\newacronym{embb}{eMBB}{enhanced Mobile Broadband}
\newacronym{en}{EN}{energy neutral}
\newacronym{end}{END}{energy neutral device}
\newacronym{enob}{ENOB}{effective number of bits}
\newacronym{eol}{EoL}{end of life}
\newacronym{epu}{EPU}{edge processing unit}
\newacronym{er}{ER}{Energy Receiver}
\newacronym{erp}{ERP}{equivalent radiated power}
\newacronym[plural=ESCs,firstplural=electronic speed controllers (ESCs)]{esc}{ESC}{electronic speed control}
\newacronym{esd}{ESD}{electrostatic discharge}
\newacronym{esl}{ESL}{electronic shelf label}
\newacronym{esr}{ESR}{equivalent series resistance}
\newacronym{et}{ET}{Energy Transmitter}
\newacronym{etsi}{ETSI}{European Telecommunications Standards Institute}
\newacronym{evd}{EVD}{eigenvalue decomposition}
\newacronym{evm}{EVM}{Error Vector Magnitude}
\newacronym{ewlb}{eWLB}{Embedded Wafer Level Ball Grid Array}
\newacronym{fa}{FA}{federation anchor}
\newacronym{fair}{FAIR}{Findability, Accessibility, Interoperability, and Reuse of digital assets}
\newacronym{fd}{FD}{front-haul distance}
\newacronym{fde}{FDE}{frequency domain equalizer}
\newacronym{fdm}{FDM}{frequency-division multiplexing}
\newacronym{fdma}{FDMA}{frequency division multiple access}
\newacronym{fem}{FEM}{finite element analysis}
\newacronym{fembb}{feMBB}{further enhanced mobile broadband}
\newacronym{fft}{FFT}{fast Fourier transform}
\newacronym{fh}{FH}{fronthaul}
\newacronym{fifo}{FIFO}{First In, First Out}
\newacronym{fim}{FIM}{Fisher information matrix}
\newacronym{fits}{FITS}{Flexible Image Transport System}
\newacronym{fom}{FoM}{figure of merit}
\newacronym{fpga}{FPGA}{field-programmable gate array}
\newacronym{fsk}{FSK}{frequency shift keying}
\newacronym{fss}{FSS}{frequency selective surface}
\newacronym{gb}{GB}{grant-based}
\newacronym{gf}{GF}{grant-free}
\newacronym{gmsk}{GMSK}{Gaussian minimum-shift keying}
\newacronym{gnb}{gNB}{Next Generation Node B}
\newacronym{gnn}{GNN}{graph neural network}
\newacronym{gnss}{GNSS}{global navigation satellite system}
\newacronym{gpclk}{GPCLK}{general purpose clock}
\newacronym{gpl}{GPL}{GNU General Public License}
\newacronym{gprs}{GPRS}{General Packet Radio Services}
\newacronym{gps}{GPS}{Global Positioning System}
\newacronym{gpu}{GPU}{graphical processing unit}
\newacronym{grc}{GRC}{GNU Radio Companion}
\newacronym{gscm}{GSCM}{geometry‐based stochastic model}
\newacronym{gsm}{GSM}{Global System for Mobile Communications}  %
\newacronym{gwp}{GWP}{Global Warming Potential}
\newacronym{harq}{HARQ}{hybrid automatic repeat request}
\newacronym{hat}{HAT}{hardware attached on top}
\newacronym{hcs}{HCS}{human-centric services}
\newacronym{hdf5}{HDF5}{Hierarchical Data Format version 5}
\newacronym{hfss}{HFSS}{High Frequency Simulator Software}
\newacronym{hmd}{HMD}{head-mounted display}
\newacronym{hpbm}{HPBM}{half power beam width}
\newacronym{HyMPRo}{HyMPRo}{Hybrid Multi-Path Routing algorithm}
\newacronym{i.i.d.}{i.i.d.}{independent and identically distributed}
\newacronym{ib}{IB}{in-band}
\newacronym{ibo}{IBO}{input back-off}
\newacronym{ic}{IC}{integrated circuit}
\newacronym{ici}{ICI}{intercarrier interference}
\newacronym{id}{ID}{information decoding}
\newacronym{idft}{IDFT}{inverse discrete Fourier transform}
\newacronym{if}{IF}{intermediate-frequency}
\newacronym{iid}{i.i.d.}{independently and identically distributed}
\newacronym{iis}{IIS}{integrated information system}
\newacronym{im}{IM}{intermodulation}
\newacronym{imu}{IMU}{inertial measurement unit}
\newacronym{io}{IO}{input/output}
\newacronym{iot}{IoT}{Internet of Things}
\newacronym{ipt}{IPT}{inductive power transfer}
\newacronym{ipy}{IPY}{Interventions per Year}
\newacronym{iq}{IQ}{in-phase and quadrature}
\newacronym{iqi}{IQI}{IQ imbalance}
\newacronym{ir}{IR}{infrared}
\newacronym{isi}{ISI}{intersymbol interference}
\newacronym{ism}{ISM}{industrial, scientific and medical}
\newacronym{isp}{ISP}{internet service provider}
\newacronym{jesd}{JESD}{Joint Electron Devices Engineering Council}
\newacronym{jfet}{JFET}{junction field effect transistor}
\newacronym{kpi}{KPI}{key performance indicator}
\newacronym{kvi}{KVI}{key value indicator}
\newacronym{larva}{LARVA}{LARge Virtual Array}
\newacronym{lca}{LCA}{life cycle assessment}
\newacronym{lco}{LCO}{lithium cobalt oxide}
\newacronym{ldo}{LDO}{Low-dropout voltage regulator}
\newacronym{ldpc}{LDPC}{low-density parity-check}
\newacronym{less}{LESS}{Low Energy Scheduler Solution}
\newacronym{lfp}{LFP}{lithium iron phosphate}
\newacronym{lic}{LIC}{lithium-ion capacitor}
\newacronym{lidar}{LiDAR}{light detection and ranging}
\newacronym{liion}{Li-ion}{lithium-ion}
\newacronym{lipo}{LiPo}{lithium polymer}
\newacronym{lis}{LIS}{large intelligent surface}
\newacronym{llh}{LLH}{log-likelihood}
\newacronym{lmmse}{LMMSE}{least minimum mean square error}
\newacronym{lmo}{LMO}{lithium ion manganese oxide}
\newacronym{lna}{LNA}{low-noise amplifier}
\newacronym{lo}{LO}{local oscillator}
\newacronym{lora}{LoRa}{long range}
\newacronym{lorawan}{LoRaWAN}{long-range wide-area network}
\newacronym{los}{LoS}{line-of-sight}
\newacronym{lp}{LP}{linear programming}
\newacronym{lpt}{LPT}{laser power transfer}
\newacronym{lpwa}{LPWA}{Low Power Wide Area}
\newacronym{lpwan}{LPWAN}{low-power wide-area network}
\newacronym{lqi}{LQI}{link quality indicator}
\newacronym{lrelu}{LReLU}{leaky rectified linear unit}
\newacronym{lrt}{LRT}{likelihood-ratio test}
\newacronym{ls}{LS}{least squares}
\newacronym{lsa}{LSA}{large synthetic array}
\newacronym{lsf}{LSF}{large-scale fading}
\newacronym{lsfc}{LSFC}{large-scale fading component}
\newacronym{ltc}{LTC}{lithium thionyl chloride}
\newacronym{lte}{LTE}{Long Term Evolution}
\newacronym{lti}{LTI}{linear time-invariant}
\newacronym{lto}{LTO}{lithium titanate}
\newacronym{m2m}{M2M}{machine to machine}
\newacronym{mac}{MAC}{Medium Access Control}
\newacronym{mate}{MATE}{millimeter-wave MIMO testbed}
\newacronym{mc}{MC}{Monte Carlo}
\newacronym{mcl}{MCL}{Maximum Coupling Loss}
\newacronym{mcs}{MCS}{modulation and coding scheme}
\newacronym{mcu}{MCU}{Microcontroller Unit}
\newacronym{mec}{MEC}{multi-access edge computing}
\newacronym{mems}{MEMS}{micro-electromechanical systems}
\newacronym{mf}{MF}{matched filter}
\newacronym{mimo}{MIMO}{multiple-input multiple-output}
\newacronym{miso}{MISO}{multiple-input single-output}
\newacronym{ml}{ML}{machine learning}
\newacronym{mlp}{MLP}{multilayer perceptron}
\newacronym{mmimo}{mMIMO}{massive MIMO}
\newacronym{mmse}{MMSE}{minimum mean square error}
\newacronym{mmtc}{mMTC}{massive machine-typed communication}
\newacronym{mmwave}{mmWave}{millimeter wave}
\newacronym{mn}{MN}{matching network}
\newacronym{mosfet}{MOSFET}{metal-oxide semiconductor field effect transistor}
\newacronym{mpc}{MPC}{multipath component}
\newacronym{mppt}{MPPT}{maximum power point tracking}
\newacronym{mr}{MR}{maximum ratio}
\newacronym{mrc}{MRC}{maximum ratio combining}
\newacronym{mrc_em}{MRC}{maximum ratio combining}
\newacronym{mrc_EM}{MRC}{Magnetic Resonance Coupling}
\newacronym{mrt}{MRT}{maximum ratio transmission}
\newacronym{mse}{MSE}{mean square error}
\newacronym{mtc}{MTC}{Machine-Type Communication}
\newacronym{multi-rat}{Multi-RAT}{multiple radio access technology}
\newacronym{music}{MUSIC}{MUltiple SIgnal Classification}
\newacronym{nb}{NB}{narrowband}
\newacronym{nbiot}{NB-IoT}{narrowband IoT}
\newacronym{nca}{NCA}{nickel cobalt aluminum}
\newacronym{netcdf}{NetCDF}{Network Common Data Form}
\newacronym{nfc}{NFC}{near-field communication}
\newacronym{ni}{NI}{National Instruments}
\newacronym{nicd}{NiCd}{nikkel cadmium}
\newacronym{nimh}{NiMH}{nikkel metal hydride}
\newacronym{nlos}{NLoS}{non-line-of-sight}
\newacronym{nmc}{NMC}{nickel manganese cobalt}
\newacronym{nmos}{nMOS}{n-channel metal-oxide semiconductor}
\newacronym{nn}{NN}{neural network}
\newacronym{nnls}{NNLS}{non-negative least squares}
\newacronym{noma}{NOMA}{non-orthogonal multiple access}
\newacronym{np}{NP}{Neyman-Pearson}
\newacronym{npbch}{NPBCH}{Narrowband Physical Broadcast Channel}
\newacronym{npss}{NPSS}{Narrow Band Primay Synchronization Signal}
\newacronym{nr}{NR}{New Radio}
\newacronym{nrs}{NRS}{Narrow Band Reference Signal}
\newacronym{nsss}{NSSS}{Narrowband Secondary Synchronization Signal}
\newacronym{ntp}{NTP}{network time protocol}
\newacronym{oai}{OAI}{OpenAirInterface} %
\newacronym{obw}{OBW}{occupied bandwidth}
\newacronym{ofdm}{OFDM}{orthogonal frequency-division multiplexing}
\newacronym{ofdma}{OFDMA}{orthogonal frequency-division multiple access}
\newacronym{ofdmim}{OFDM-IM}{OFDM with index modulation}
\newacronym{oob}{OOB}{out-of-band}
\newacronym{ook}{OOK}{on-off keying}
\newacronym{oran}{O-RAN}{open radio-access network}
\newacronym{os}{OS}{operating system}
\newacronym{ota}{OTA}{over-the-air}
\newacronym{otaa}{OTAA}{over-the-air authentication}
\newacronym{p1}{P1}{Phase 1}
\newacronym{p2}{P2}{Phase 2}
\newacronym{p2p}{P2P}{point-to-point}
\newacronym{pa}{PA}{power amplifier}
\newacronym{pae}{PAE}{power-added efficiency}
\newacronym{pana}{PanA}{Panel A}
\newacronym{panb}{PanB}{Panel B}
\newacronym{papr}{PAPR}{peak-to-average power ratio}
\newacronym{pc}{PC}{pilot count}
\newacronym{pcb}{PCB}{printed circuit board}
\newacronym{pcg}{PCG}{power consumption gain}
\newacronym{pcie}{PCIe}{Peripheral Component Interconnect Express}
\newacronym{pcsi}{PCSI}{perfect channel state information}
\newacronym{pd}{PD}{powered device}
\newacronym{pdcch}{PDCCH}{physical downlink control channel}
\newacronym{pdf}{PDF}{probability density function}
\newacronym{pdp}{PDP}{power delay profile}
\newacronym{pdsch}{PDSCH}{physical downlink shared channel}
\newacronym{peb}{PEB}{positioning error bound}
\newacronym{per}{PER}{packet error rate}
\newacronym{pg}{PG}{path gain}
\newacronym{pgd}{PGD}{proximal gradient descent}
\newacronym{phy}{PHY}{physical}
\newacronym{pl}{PL}{path loss}
\newacronym{pll}{PLL}{phase-locked loop}
\newacronym{pmf}{PMF}{polymer microwave fiber}
\newacronym{poe}{PoE}{power-over-Ethernet}
\newacronym{pps}{1PPS}{pulse per second}
\newacronym{prbs}{PRBs}{Physical Resource Blocks}
\newacronym{ps}{PS}{Processing System}
\newacronym{psd}{PSD}{power spectral density}
\newacronym{pse}{PSE}{power sourcing equipment}
\newacronym{psk}{PSK}{phase shift keying}
\newacronym{psm}{PSM}{power saving mode}
\newacronym{pss}{PSS}{primary synchronisation signal}
\newacronym{ptp}{PTP}{precision-time protocol}
\newacronym{ptrs}{PTRS}{Phase-Tracking Reference Signals}
\newacronym{ptw}{PTW}{paging time window}
\newacronym{pw}{PW}{planar wavefront}
\newacronym{pwm}{PWM}{pulse width modulation}
\newacronym{qam}{QAM}{quadrature amplitude modulation}
\newacronym{qos}{QoS}{quality-of-service}
\newacronym{qpsk}{QPSK}{quadrature phase-shift keying}
\newacronym{quadriga}{QuaDRiGa}{QUAsi Deterministic RadIo channel GenerAtor}
\newacronym{ra}{RA}{Random Access}
\newacronym{ram}{RAM}{random-access memory}
\newacronym{ran}{RAN}{radio access network}
\newacronym{rar}{RAR}{Random Access Response}
\newacronym{rat}{RAT}{radio access technology}
\newacronym{rb}{Rb}{Rubidium}
\newacronym{rbs}{RBS}{radio base station}
\newacronym{rbw}{RBW}{resolution bandwidth}
\newacronym{rcs}{RCS}{radar cross section}
\newacronym{re}{RE}{radio element}
\newacronym{relu}{ReLU}{rectified linear unit}
\newacronym{rf}{RF}{radio frequency}
\newacronym{rfeh}{RFEH}{radio frequency energy harvesting}
\newacronym{rfic}{RFIC}{radio-frequency integrated circuit}
\newacronym{rfid}{RFID}{radio frequency identification}
\newacronym{rfpt}{RFPT}{radio frequency power transfer}
\newacronym{rfsoc}{RFSoC}{Radio Frequency System-on-Chip}
\newacronym{rir}{RIR}{room impulse response}
\newacronym{ris}{RIS}{reflective intelligent surface}%
\newacronym{rllmtc}{RLLMTC}{reliable low latency machine type communication}
\newacronym{rls}{RLS}{recursive least squares}
\newacronym{rms}{RMS}{root-mean-square}
\newacronym{rmse}{RMSE}{root-mean-square error}
\newacronym{rmt}{RMT}{random matrix theory}
\newacronym{rof}{RoF}{radio-over-fiber}
\newacronym{ros}{ROS}{robot operating system}
\newacronym{rpi}{RPi}{Raspberry Pi}
\newacronym{rrc}{RRC}{Radio Resource Connection}
\newacronym{rreq}{RREQ}{route request packet}
\newacronym{rsrp}{RSRP}{Reference Signals Received Power}
\newacronym{rss}{RSS}{received signal strength}
\newacronym{rssi}{RSSI}{received signal strength indicator}
\newacronym{rtc}{RTC}{real time clock}
\newacronym{rtk}{RTK}{real time kinematics}
\newacronym{rts}{RTS}{ray tracing simulator}
\newacronym{ru}{RU}{Radio Unit}
\newacronym{rv}{RV}{random variable}
\newacronym{rw}{RW}{RadioWeaves}
\newacronym{rx}{RX}{receiver}
\newacronym{rzf}{RZF}{regularized zero forcing}
\newacronym{s-parameter}{S-parameter}{scattering parameter}
\newacronym{sa}{SA}{synchronization anchor}
\newacronym{sar}{SAR}{specific absorption rate}
\newacronym{sbl}{SBL}{sparse Bayesian learning}
\newacronym{scfdma}{SCFDMA}{single-carrier frequency division multiple access}
\newacronym{sdg}{SDG}{Sustainable Development Goal}
\newacronym{sdm}{SDM}{sigma-delta modulator}
\newacronym{sdn}{SDN}{software-defined network}
\newacronym{sdof}{SDoF}{sigma-delta over fiber}
\newacronym{sdr}{SDR}{software-defined radio}
\newacronym{sei}{SEI}{specific emitter identification}
\newacronym{ser}{SER}{symbol-error rate}
\newacronym{sf}{SF}{spreading factor}
\newacronym{sfn}{SFN}{single frequency network}
\newacronym{sfp}{SFP}{small form-factor pluggable}
\newacronym{SigMF}{SigMF}{Signal Metadata Format}
\newacronym{simo}{SIMO}{single-input multiple-output}
\newacronym{sinr}{SINR}{signal-to-interference-plus-noise ratio}
\newacronym{siso}{SISO}{single-input single-output}
\newacronym{slam}{SLAM}{simultaneous localization and mapping}
\newacronym{slc}{SLC}{spatial leakage suppression}
\newacronym{slerp}{SLERP}{spherical linear interpolation}
\newacronym{sma}{SMA}{SubMiniature version A}
\newacronym{smc}{SMC}{specular multipath component}
\newacronym{smps}{SMPS}{switched mode power supply}
\newacronym{sndr}{SNDR}{signal-to-noise-and-distortion ratio}
\newacronym{snidr}{SNIDR}{signal-to-noise-and-interference-and-distortion ratio}
\newacronym{snr}{SNR}{signal-to-noise ratio}
\newacronym{soc}{SoC}{state of charge}
\newacronym{sram}{SRAM}{static random-access memory}
\newacronym{srd}{SRD}{short-range device}
\newacronym{srs}{SRS}{Sounding Reference Signal}
\newacronym{ssb}{SSB}{synchronisation signal block}
\newacronym{ssd}{SSD}{solid state drive}
\newacronym{ssq}{SSQ}{simulator sickness questionnaire}
\newacronym{steam}{STEAM}{science, technology, engineering, the arts, and mathematics}
\newacronym{svd}{SVD}{singular value decomposition}
\newacronym{sw}{SW}{spherical wavefront}
\newacronym{synce}{SyncE}{Synchronous Ethernet}
\newacronym{tau}{TAU}{tracking area update}
\newacronym{tcer}{TCER}{transported to consumed energy ratio}
\newacronym{tcp}{TCP}{Transmission Control Protocol}
\newacronym{tdd}{TDD}{time division duplexing}
\newacronym{tdoa}{TDOA}{time-difference-of-arrival}
\newacronym{thz}{THz}{Terahertz}
\newacronym{to}{TO}{timing offset}
\newacronym{toa}{TOA}{time-of-arrival}
\newacronym{tof}{ToF}{time-of-flight}
\newacronym{tosm}{TOSM}{through-open-short-match}
\newacronym{trl}{TRL}{technology readyness level}
\newacronym{trp}{TRP}{Transmission Reception Point}
\newacronym{tsn}{TSN}{time-sensitive networking}
\newacronym{ttm}{TTM}{time to market}
\newacronym{ttn}{TTN}{The Things Network}
\newacronym{tx}{TX}{transmitter}
\newacronym{uav}{UAV}{unmanned aerial vehicle}
\newacronym{ucie}{UCIe}{Universal Chiplet Interconnect Express}
\newacronym{udp}{UDP}{User Datagram Protocol}
\newacronym{ue}{UE}{user equipment}
\newacronym{ugv}{UGV}{unmanned ground vehicle}
\newacronym{uhd}{UHD}{USRP hardware driver}
\newacronym{uhf}{UHF}{ultra-high frequency}
\newacronym{ul}{UL}{uplink}
\newacronym{ula}{ULA}{uniform linear array}
\newacronym{ummtc}{umMTC}{ultra massive machine type communication}
\newacronym{un}{UN}{United Nations}
\newacronym{upa}{UPA}{uniform planar array}
\newacronym{ura}{URA}{uniform rectangular array}
\newacronym{urllc}{URLLC}{ultra-reliable low-latency communications}
\newacronym{usrp}{USRP}{universal software radio peripheral}
\newacronym{uv}{UV}{unmanned vehicle}
\newacronym{uwb}{UWB}{ultrawideband}
\newacronym{v2v}{V2V}{vehicle-to-vehicle}
\newacronym{vep}{VEP}{virtual edge platform}
\newacronym{vlc}{VLC}{visible light communication}
\newacronym{vlp}{VLP}{visible light positioning}
\newacronym{vna}{VNA}{vector network analyzer}
\newacronym{votable}{VOTable}{Virtual Observatory Table}
\newacronym{vr}{VR}{virtual reality}
\newacronym{wb}{WB}{wideband}
\newacronym{wban}{WBAN}{wireless body area network}
\newacronym{wimax}{WiMAX}{Worldwide Interoperability for Microwave Access}
\newacronym{wlan}{WLAN}{wireless LAN}
\newacronym{wpt}{WPT}{wireless power transfer}
\newacronym{wr}{WR}{White Rabbit}
\newacronym{wrsn}{WRSN}{wireless rechargeable sensor network}
\newacronym{wsn}{WSN}{wireless sensor network}
\newacronym{xets}{XETS}{cross exponentially tapered slot}
\newacronym{xr}{XR}{extended reality}
\newacronym{z3ro}{Z3RO}{zero third-order distortion}
\newacronym{zf}{ZF}{zero-forcing}
\newacronym{zmcscg}{ZMCSCG}{zero mean circularly symmetric complex Gaussian}

\newacronym{swipt}{SWIPT}{simultaneous wireless information and power transfer}

%% file: tools.tex
\usepackage[dvipsnames,svgnames, table]{xcolor} %

\usepackage{listofitems}
\usepackage{pgffor}

\makeatletter
\def\myColorList{LimeGreen,BrickRed,Fuchsia,Bittersweet,YellowOrange, YellowGreen, WildStrawberry} %
\readlist\ColorList\myColorList

\newcommand{\defineauthors}[1]{

    \foreach \x [count=\xi from 1] in {#1} { 
    \expandafter\xdef\csname\x\endcsname####1{\noexpand\textcolor{\ColorList[\xi]}{[\unexpanded\expandafter{\x}: ####1]}}%
    }
}
\makeatother

\usepackage{array}
\newcolumntype{H}{>{\setbox0=\hbox\bgroup}c<{\egroup}@{}}

\newcounter{assumcounter}
\renewcommand{\theassumcounter}{A\arabic{assumcounter}}

\newcommand{\newassumption}[1]{%
    \refstepcounter{assumcounter}%
    \label{a:#1}%
    (\theassumcounter)%
}

%% file: sections/00_Introduction.tex
\glsresetall
\section{Introduction}

In recent years, there has been a growing body of literature focusing on \gls{wpt}, covering both \gls{rf} and hardware solutions~\cite{Huang2019Wireless, Wagih2020Antennas, VanMulders2022}, as well as communication, signal, and system designs~\cite{Zeng2017, clerckx2021wireless}. Despite the significant efforts dedicated to the engineering of energy harvesters, as highlighted in~\cite{clerckx2018MicrowaveMagazine} and illustrated in~\cite{Valenta2014Microwave}, there has been comparatively less emphasis on the co-design of signal and circuit-level, taking into account inherent nonlinear characteristics of \gls{wpt} systems.

Several notable studies have explored different aspects of \gls{wpt} optimization. For instance,~\cite{7867826} investigates the Pareto boundary, aiming for weighted-sum-power maximization (WSPMax) under fixed total transmit power constraints. Similarly,~\cite{9411899} focuses on increasing the output DC power by jointly optimizing multi-sine waveforms and beamforming, accounting for rectenna nonlinearity. The work presented in~\cite{10136750} delves into waveform design considering non-linear high-power amplifiers (HPA) and energy harvester (EH) non-linearities. Additionally, an overview of \gls{swipt} is presented in~\cite{8476597}.

Contrary to prior research, we recognize that \glspl{er} have distinct energy needs. For instance, \glspl{esl} may only require intermittent screen updates, eliminating the need for a continuous \gls{dc} power supply. As a result, instead of maximizing the received \gls{dc} power for all \glspl{er}, our approach prioritizes delivering the essential energy to each \gls{er} with the least system transmit power. 

\begin{figure}
    \centering
    \includegraphics[width=0.98\linewidth]{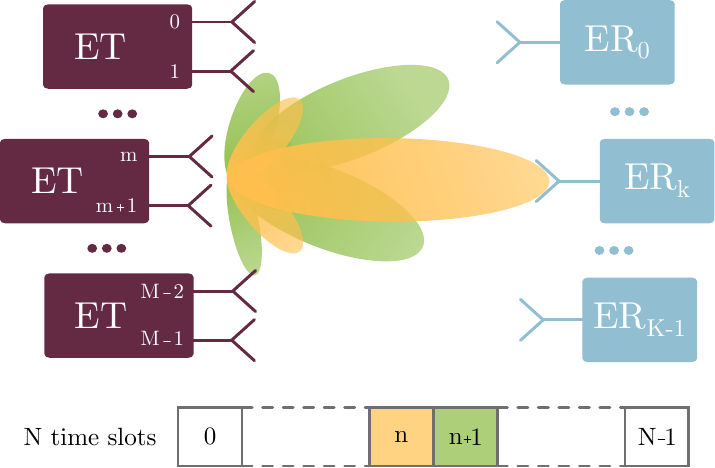}
    \caption{Illustration of a distributed \gls{wpt} system, where several \glspl{et} transmit signals over \(M\) antennas to \(K\) \glspl{er} to charge them over \(N\) time slots. In this example, during slot \(n\) only \gls{er} $k$ is targeted, while at \(n+1\) $\text{ER}_0$ and $\text{ER}_{K-1}$ is being charged.
    }\label{fig:system}
\end{figure}

We begin by introducing the system model, encompassing the \gls{rf}-to-\gls{dc} conversion model and the underlying assumptions of this study. In \cref{sec:algo}, we formalize the problem for both non-coherent and coherent systems. The specific scenario under consideration is detailed in \cref{sec:eval-scenario}. We then proceed to evaluate the problem optimization through numerical and experimental assessments in \cref{sec:eval}. Finally, we conclude with a summary and offer insights into future research directions in \cref{sec:concl}.

%% file: sections/02_SignalProcessing.tex
\section{System Model}\label{sec:system-model}
Consider a multi-antenna system as illustrated in~\cref{fig:system}, where \(M\) antennas are available to wirelessly charge \(K\) \glspl{er}. In the system model, we do not impose how these antennas are distributed, i.e., distributed or co-located. The received \gls{rf} power \(P_\text{RF}\) is converted to \gls{dc} power \(P_\text{DC}\) by the harvester in the \gls{er}. The utilized \gls{rf}-to-\gls{dc}  model is,
\begin{equation}
    P_\text{DC} = \alpha P_\text{RF} - \beta,\label{eq:harvester}
\end{equation}
where $\alpha$ and  $\beta$ are real positive coefficient, and $\alpha$ is the \gls{rf} harvester efficiency and $\beta$ is the minimum received \gls{rf} power required to provide a stable voltage level to the \gls{er} circuitry. See~\cref{sec:esl-char} for more details regarding the RF-to-DC power behavior.

Over \(N\) time slots, each lasting \(T\) seconds, the objective is to harvest a certain amount of energy \(E\) at each \gls{er}. The \(M\) antennas can be hosted on one (co-located system) or several (distributed system) \glspl{et}. An illustration of a distributed system with two antennas per \gls{et} is shown in~\cref{fig:system}. In this work, two scenarios are investigated, i.e., a non-coherent and a coherent scenario. In both scenarios, each antenna transmits the same narrowband signal. As opposed to communication signals, this signal carries only energy, not information. Hence, the transmitted signal is the same for all \glspl{er}.

\textit{Assumptions.} 
Below is a list of assumptions used throughout the manuscript.

\textbf{\newassumption{duty-cycle}} Although specified in some frequency bands, here no duty cycle limits are assumed during transmission. 
\textbf{\newassumption{pa-lin}} The \glspl{pa} are presupposed to work in the linear regime. This can be obtained by employing a sufficient back-off or using techniques such as \gls{dpd}. 
\textbf{\newassumption{static-csi}}  The channel is considered static over the $NT$ time period. Based on the findings~\cite[Fig.~1]{10118929}, this can be considered reasonable when both the \glspl{et} and \glspl{er} are static.

\textit{Specific to the non-coherent system}: \textbf{\newassumption{iid-phase}} An \gls{iid} phase difference is assumed for all received signals. This can be assumed when having different \glspl{mpc} and, additionally, the transmit antennas each use different transmit phases. Additionally, \textbf{\newassumption{perfect-gain}} the channel gain, i.e., \gls{pl}, is perfectly known.

\textit{Specific to the coherent system}: \textbf{\newassumption{perfect-csi}} perfect \gls{csi} and \textbf{\newassumption{perfect-sync}} perfect synchronization is assumed.

\section{Optimal Distributed Precoding and Sweeping}\label{sec:algo}
Below, we formalize the problem of charging \(K\) \esls with \(M\) antennas in two scenarios. The symbols used are summarized in~\cref{tab:symbols}.\footnote{The mathematical notations are detailed in \url{https://github.com/wavecore-research/math-notations}}.

\input{sections/02a_partial_CSI}

\input{sections/02b_full_CSI}

%% file: sections/02a_partial_CSI.tex
\subsection{Non-Coherent System}

Under (\ref{a:iid-phase}), on average, the received \gls{rf} power is equal to the sum of all powers. Hence, using~\cref{eq:harvester}, the harvested \gls{dc} power $\tilde{p}_{k,n}$ at \esl $k$ at time slot $n$ is \textit{on average}\footnote{Here, by average we mean the expectation computed with respect to the \gls{iid} phase distribution.},
\begin{equation*}
    \tilde{p}_{k,n} =  \alpha \sum_m^M \left( |h_{m,k}|^2p_{m,n} \right) - \beta,
\end{equation*}
where $p_{m,n}$ is the transmit power of antenna $m$ at time slot $n$ and $h_{m,k}$ the complex channel coefficient between antenna $m$ and \esl $k$, which are known (\ref{a:perfect-gain}). 
To minimize the total transmit energy while providing the required \gls{dc} energy, the problem can be written as,

\begin{align}
	\min_{p_{m,n}} \sum_{n}^{N}\sum_{m}^{M} T p_{m,n} \text{ s.t. }   T \sum_{n}^{N}  \tilde{p}_{k,n} &\geq E,\ \forall k \\
    p_{m,n} &\leq P_\text{max},\ \forall m,n,\label{eq:non-coherent-problem}
\end{align}
with as constrain that the received energy over all time slots should be at least $E$ for each \esl and the transmit power does not exceed the maximum power constraint.

%% file: sections/02b_full_CSI.tex
\subsection{Coherent System}
In the fully coherent scenario, perfect \gls{csi} (\ref{a:perfect-csi}) and perfect synchronization (\ref{a:perfect-sync}) is assumed.  
Each antenna precodes the narrowband transmit signal by a beamformer weight. At each time slot $n$, the beamformer can be reconfigured to preferably send to a given (or set of given) ER(s), as illustrated in~\cref{fig:system}. Hence, the beamformer vector is assumed to depend on the time slot. The beamforming vector at time slot $n$ is denoted by $\vect{w}_n \in \mathbb{C}^{M\times 1}$. Assuming the transmit signal has unit power\footnote{This in contrast to the non-coherent scenario, where the transmit power is not unit power, but $p_{m,n}$.}, the total transmit power at time slot $n$ is $\|\vect{w}_n\|^2$ constraint to $P_\text{max}$. As a cost function for designing the beamformer weights, we would like to minimize the total transmit energy over the $n$ time slots
\begin{align*}
	\min_{\vect{w}_0,\hdots,\vect{w}_{N}} T\sum_{n=0}^{N-1}\|\vect{w}_n\|^2,
\end{align*}
with as constraint, that the received energy over all time slots should be at least $E$ for each \esl.  
The received \gls{rf} signal power at time slot $n$ for \esl \(k\) is,
\begin{align}
    p_\text{RF}{}_{k,n} = |\vect{h}_k^T\vect{w}_n|^2.
\end{align}

Taking into the account the efficiency of the harvester~\cref{eq:harvester}, the optimization problem becomes,
\begin{align*}
	\min_{\vect{w}_0,\hdots,\vect{w}_{N-1}} \sum_{n=0}^{N-1}T\|\vect{w}_n\|^2\\ \text{ s.t. } \sum_{n=0}^{N-1} T \alpha |\vect{h}_k^T\vect{w}_n|^2 \geq E + NT\beta,\ \forall k.
\end{align*}

The cost function is convex (quadratic) but the inequality is not convex because of the sign of the inequality. Therefore, we can rewrite the constraint using
\begin{align*}
	\sum_{n=0}^{N-1} T\alpha |\vect{h}_k^T\vect{w}_n|^2 &= \sum_{n=0}^{N-1} T \alpha \vect{h}_k^T \vect{w}_n \vect{w}_n^H   \vect{h}_k^*\\
	&= \tr\left[\mat{G}_k\mat{X}\right]  
\end{align*}
where we defined $\mat{G}_k=\vect{h}_k^*\vect{h}_k^T$ and $\mat{X} = \alpha T \sum_{n=0}^{N-1} \vect{w}_n \vect{w}_n^H \in \mathbb{C}^{M\times M}$. By its definition, $\mat{X}$ is positive semidefinite and has a rank less or equal to $\min(M,N)$. In practice, we can reasonably expect that, by design, a large number of time slots will be available so that $N\geq M$ and hence, the rank of $\mat{X}$ is only constrained by its dimension $M$. This makes sense, we would like not to constraint the optimization and have $N$ at least as large as $K$ to be able to sweep across each \esl. If not useful, the time slots will not be used by the optimization and the precoding weights at those timeslots will be set to zero. We can also note that $\tr[\mat{X}]=\sum_{n=0}^{N-1}\alpha T\|\vect{w}_n\|^2$. Hence, the optimization problem can be rewritten as a function of $\mat{X}$
\begin{align}
	\min_{\mat{X}  \succeq  0} \tr[\mat{X}]  \text{ s.t. } \tr\left[\mat{G}_k\mat{X}\right]  \geq E + NT\beta,\ \forall k.\label{eq:coherent-problem}
\end{align}
Given that $N\geq M$, this already reduces the dimension of the optimization variables from $NM$ to $M^2$. Moreover, the cost function and constraint have a linear form, which is convex. This together with positive semidefinite constraint $\mat{X}  \succeq  0$, yields a standard semidefinite programming structure and can be optimized using conventional solvers. 

Based on the optimal $\mat{X}$, we can find back $\vect{w}_n$ using, e.g., the \gls{evd}.
Consider that \( \mat{Q} \) is a unitary matrix \( \in \mathbb{C}^{M \times M} \) containing eigenvectors of \( \mat{X} \) and  \( \mat{\Lambda} \) is a diagonal matrix of size \( \in \mathbb{C}^{M \times M} \) containing eigenvalues of \( \mat{X} \). Then, the precoding matrix $\mat{W}$ can be found as,

\begin{align*}
\mat{X} &=\alpha T \mat{W}\mat{W}^H\\
    &=\mat{Q} \mat{\Lambda} \mat{Q}^H\\
\mat{W} &= \frac{1}{\sqrt{\alpha T}} \mat{Q} \mat{\Lambda}^{1/2},
\end{align*}

where \( \mat{\Lambda}^{1/2} \) denotes taking the square root of each diagonal element of \(  \mat{\Lambda} \).  Only the $S$ columns of $\mat{W}$ are used, associated with the $S$ non-zero eigenvalues. %
This can be interpreted as using only $S$ out of $N$ timeslots.

%% file: sections/03_Evaluation.tex
\section{Scenario Description for Evaluation}
The proposed optimal distributed precoding and power allocation scheme is evaluated based on a representative use case, i.e., a wirelessly charging \acrfullpl{esl} in a grocery store. The system is evaluated both numerically and experimentally, where the used parameters and their default values are summarized in~\cref{tab:symbols}. Unless stated otherwise, the default values are used.

\begin{table}[h]
    \centering
    \caption{Used parameters, symbols, and expressions with their default value and units.}\label{tab:symbols}
    \begin{tabular}{l l l l}
    \toprule
    Parameter & Symbol & Default value& Unit\\
    \midrule
    Carrier frequency & $f_c$ & \num{0.917} &\si{\giga\hertz}\\
    Required energy per node & $E_k$ & \num{500} & \si{\milli\joule}\\
    Number of time slots & $N$ & $(1,\hdots,N=M)$\\
    Number of antennas &$M$ & $(1,\hdots,84)$\\
    Time per timeslot & $T$ & -& \si{\second}\\
    Time period & $N T$ & \num{12} & \si{\hour}\\
    Number of \glspl{esl} & $K$ & 240 \\
    RF-to-DC efficiency & $\alpha$ & 0.16 & -\\
    Min. req. received RF power & $\beta$ &0.0158& \si{\milli\watt}\\
    Max. power per antenna & $P_\text{max}$ & \num{4000} & \si{\milli\watt}\\
    \bottomrule
    \end{tabular}
\end{table}

\subsection{Scenario}\label{sec:eval-scenario}
\begin{figure}[h]
    \centering
    \includegraphics[width=0.98\linewidth]{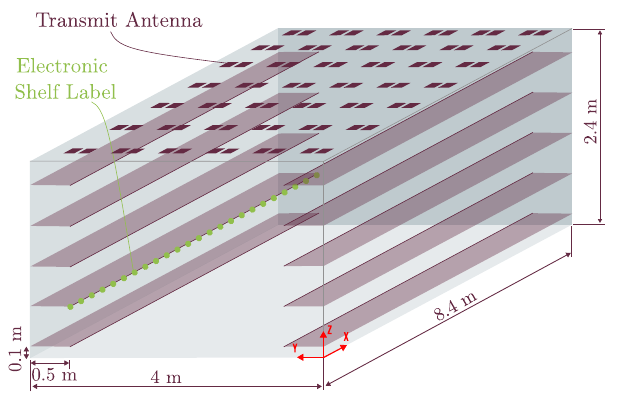}
    \caption{Illustration of the Techtile testbed~\cite{CallebautTechtile} simulating a supermarket aisle, featuring 84 ceiling-mounted antennas. The green circular areas indicate the placement of \glspl{esl} on a single shelf.}\label{fig:techtile-setup}
\end{figure}

For both numerical analysis and measurements, we focus on a single shopping aisle within a grocery store as the deployment scenario. A representation of this setup is shown in~\cref{fig:techtile-setup}. In this configuration, two shopping racks flank a \SI{4}{\meter} wide aisle, each rack extending \SI{8.4}{\meter} in length. Five shelves, spaced evenly on the \SI{2.1}{\meter} tall racks, hold the \glspl{esl} at a depth of \SI{0.5}{\meter}. We employ \(K=84\) transmit antennas to power \(M=240\) \glspl{esl}. These transmit antennas are integrated into the ceiling, providing power to the \glspl{esl}. The antenna array comprises pairs of antennas evenly spread across the ceiling. The \num{84} antennas are grouped into sets of \num{2}, consistent with the testbed described in \cref{sec:eval-exp}.

\subsection{ESL Characterization}\label{sec:esl-char}
\input{sections/01_ESL}

\subsection{Channel Model}
We utilize the \gls{los} \textit{indoor factory} \gls{pl} model of ETSI and 3GPP~\cite{ETSImodels}, which is suitable for environments with abundant metal, akin to the supermarket under consideration. 
We assume that the large-scale fading $L_{m,k}$ does not change over the 12-hour window (\ref{a:static-csi}) and is known at the antenna array (\ref{a:perfect-gain}). 
The complex channel can then be modeled as $h_{m,k}= \sqrt{L_{m,k}} e^{-\jmath \phi_{m,k}}$.
The real positive coefficient $L_{m,k}$ models the \gls{pl} and is defined in~\cite{ETSImodels}. A (uniform) random angle $\phi_{m,k}$ is considered. 
Also, here, the channel is considered constant (\ref{a:static-csi}) and known to the transmit antennas (\ref{a:perfect-csi}).

\subsection{Antenna Selection Procedure}
\begin{figure}
    \centering
    \input{figures/techtile/clusters-6}
    \caption{Illustration of antenna selection method for different number of antennas. Dark dots represent the selected antennas, while the six pentagon markers indicate the centers of six clusters. The clusters are indicated by having the same colors. The clusters are determined by k-means clustering. The positions of non-selected antennas are shown with reduced transparency.}\label{fig:antenna-selection}
\end{figure}
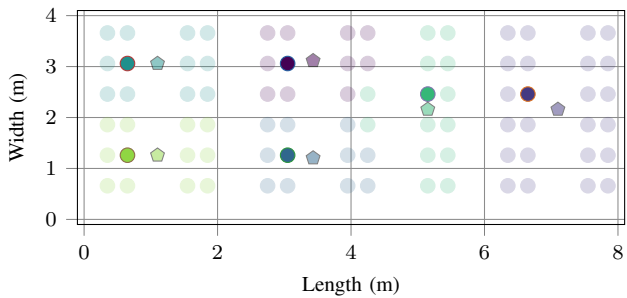

In the evaluation, the impact of increasing the number of antennas in the system is investigated. In order to have a fair comparison of the performance metrics for different number of antennas, the goal is to select a subset of antennas which are as evenly distributed as possible in the area (ceiling). To accomplish this, we designate a subset \(M_s\) (\(M_s=0,\ldots,|\mathcal{M}|\)) from the available antennas \(\mathcal{M}\) using \mbox{k-means} clustering. We form \(M_s\) clusters, and then choose the antenna nearest to the center of each cluster. This selection process is depicted in~\cref{fig:antenna-selection}.

\section{Numerical and Experimental Evaluation}\label{sec:eval}

The optimization problems, \cref{eq:non-coherent-problem} and \cref{eq:coherent-problem}, are solved using MOSEK and CVXPY in Python. The scripts are published in open-source, including additional figures~\cite{GithubProject}~\faicon{github}.

\subsection{Numerical Evaluation}
The system's performance is assessed on its ability to deliver adequate energy to the devices without exceeding the required amount, aiming for precise energy delivery. Additionally, we analyze the total used average transmit power. This evaluation considers varying numbers of antenna elements and different system configurations (non-coherent vs. coherent).

\subsubsection{Impact on increasing the number of antennas}
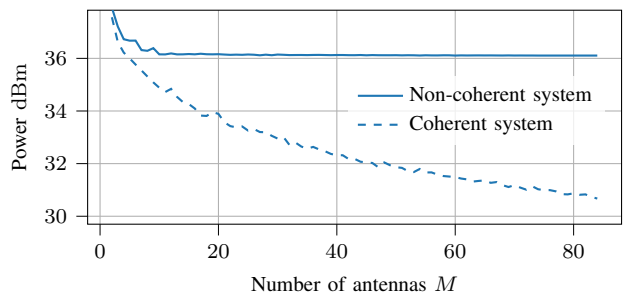
\begin{figure}[h]
    \centering
    \input{figures/tx-powers-per-M.tex}
    \caption{Total transmit power of all deployed antennas for the coherent and non-coherent system. Initially, with a limited number of antennas, both systems exhibit similar transmit power, when the number of antennas increases, the non-coherent system plateaus, whereas the coherent system continues to improve.}\label{fig:tx-powers}
\end{figure}

\cref{fig:tx-powers} illustrates the required total transmit power of the entire system relative to the number of antennas utilized.\footnote{The power values should be interpreted as if the system were transmitting continuously at that level over the full \SI{12}{\hour} period, although this is not the actual case, see~\cref{fig:timeslots}.}
It can be observed that the performance of the non-coherent system plateaus quickly, whereas the coherent system shows improvement with an increasing number of antennas. The performance of the non-coherent system levels off around \num{10} antennas, with the optimization consistently utilizing only approximately \num{10} antennas regardless of the total number available. This in contrast to the coherent system, which utilizes all available antennas. \textbf{The optimization problem favors sparsity in the spatial (antenna) domain in the non-coherent case, which is not the case in the coherent system, where all antennas are utilized.}

\subsubsection{Distribution of harvested energy} 
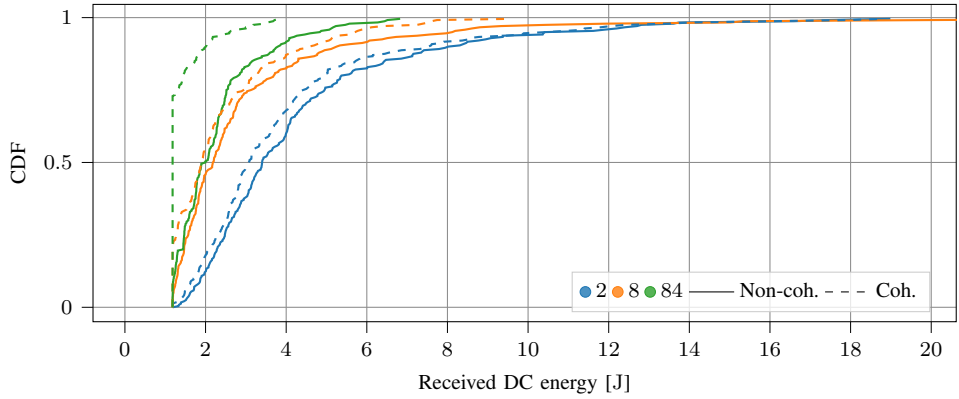
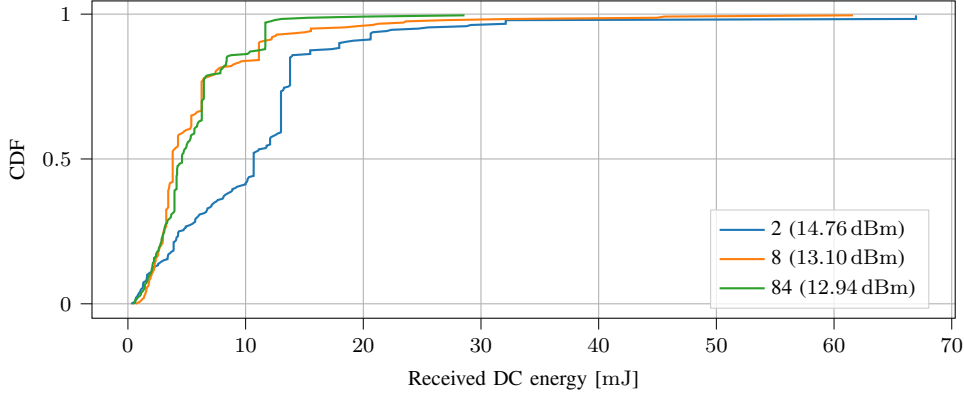
\begin{figure*}[h]
     \centering
     \begin{subfigure}[t]{0.8\textwidth}
      \centering
    \input{figures/techtile/rx-cdf-total}
    \caption{Numerical evaluation. \Gls{cdf} in the non-coherent (solid) and coherent (dashed) scenario.}\label{fig:cdf-dc-rx-sim}
    \end{subfigure}%
     \\
     \begin{subfigure}[t]{0.8\textwidth}
         \centering
         \input{figures/techtile/non-coherent/measurements}
        \caption{Experimental evaluation. Non-coherent measurements including the used total transmit power in brackets.}\label{fig:cdf-dc-rx-meas}
     \end{subfigure}
     \caption{Simulation (a) and measurement-based (b) \gls{cdf} of the received \gls{dc} energy over the full 12-hour window for 2, 8 and 84 antennas.}\label{fig:cdf-dc-rx}
\end{figure*}

In both systems, the accuracy of the targeted received \gls{dc} power improves with having more antennas, as indicated  by the \gls{cdf} in~\cref{fig:cdf-dc-rx-sim}. With increased number of antennas, i) the total transmit power is decreased, ii) the variance of received energy over all \glspl{esl} is decreased and iii) the maximum received \gls{dc} energy is reduced. The coherent system outperforms the non-coherent system in aforementioned regards, while at the same time uses less transmit power in~\cref{fig:cdf-dc-rx-sim}.

\subsubsection{Effect of number of timeslots}
\begin{figure}[h]
    \centering
    \input{figures/techtile/coherent/tx-powers-per-N}
    \caption{The used transmit powers (z-axis) over the full-time window (x-axis) for different number of available antennas (y-axis) in the coherent system.}%
    \label{fig:timeslots}
\end{figure}
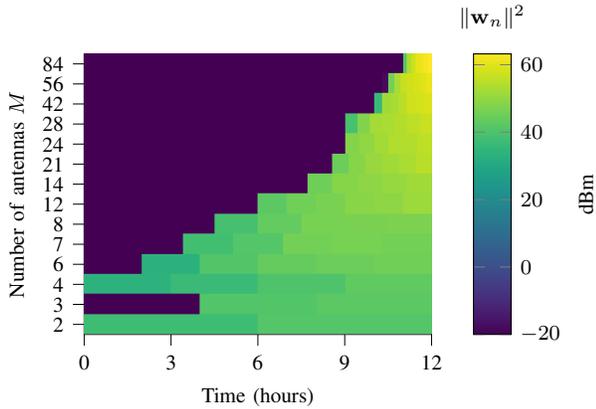

In the non-coherent system, the solution keeps the same transmit power constant per antenna for all time slots and only uses a selection of the available antennas. This in contrast to the coherent, which utilizes only a few time slots and employs all antennas, as shown in~\cref{fig:timeslots}. \textbf{In contrast to the spatial domain, in the temporal domain, the optimization problem favors sparsity in the coherent case, while using all time slots in the non-coherent system.}

%% file: sections/01_ESL.tex
The system and the considered \gls{esl} operate at \SI{917}{\mega\hertz}. In this band~\cite{ETSI_EN_302_208}, a maximum transmit power of \SI{4}{\watt} is permitted.

The considered batteryless \gls{esl} device comprises an antenna, energy harvester, energy buffer, microcontroller, and an e-paper display. RF energy is captured by the antenna and converted to DC energy by the harvester IC. Once the energy buffer reaches the required level, the microcontroller initializes, receives downlink updates, and refreshes the e-paper display. The daily DC energy requirement is pivotal for our proposed optimization, influenced by the \gls{esl}'s refresh rate and the energy needed for a screen update. The primary energy consumer in an \gls{esl} is the \textit{E-Ink Raw Display}. As indicated in~\cite{epaperscreen}, updating the display consumes \SI{450}{\milli\joule} (\SI{30}{\milli\watt} over \SI{15}{\second}), with an additional \SI{50}{\milli\joule} required for \gls{mcu} operation and demodulating downlink signals. Thus, each \gls{esl} screen update demands \SI{500}{\milli\joule}. Given two updates daily, approximately \SI{1}{\watt\second/day/\gls{esl}} is required. We consider the performance an \gls{rf} energy harvesting test IC, designed to amplify and rectify low-input power levels to DC voltages. Its conversion efficiency, \(\alpha\), and minimum required \gls{rf} input power, \(\beta\), are detailed in~\cref{tab:symbols}, serving as key parameters in both numerical and experimental evaluations.

%% file: figures/techtile/clusters-6.tex
\begin{tikzpicture}

\definecolor{crimson2143940}{RGB}{214,39,40}
\definecolor{darkcyan32144140}{RGB}{32,144,140}
\definecolor{darkgray176}{RGB}{176,176,176}
\definecolor{darkorange25512714}{RGB}{255,127,14}
\definecolor{darkslateblue48103141}{RGB}{48,103,141}
\definecolor{darkslateblue6857130}{RGB}{68,57,130}
\definecolor{forestgreen4416044}{RGB}{44,160,44}
\definecolor{indigo68184}{RGB}{68,1,84}
\definecolor{mediumpurple148103189}{RGB}{148,103,189}
\definecolor{mediumseagreen53183120}{RGB}{53,183,120}
\definecolor{sienna1408675}{RGB}{140,86,75}
\definecolor{steelblue31119180}{RGB}{31,119,180}
\definecolor{yellowgreen14421467}{RGB}{144,214,67}

\begin{axis}[
tick align=outside,
tick pos=left,
xlabel={Length (m)},
xmin=-0.1, xmax=8.1,
ylabel={Width (m)},
ymin=-0.1, ymax=4.1,
width=\linewidth,
height=0.5\linewidth,
]
\addplot [steelblue31119180, mark=*, mark size=2.73861278752583, mark options={solid,fill=indigo68184}, only marks]
table {%
3.05 3.06
};
\addplot [indigo68184, opacity=0.2, mark=*, mark size=2.73861278752583, mark options={solid}, only marks]
table {%
3.95 2.46
3.95 3.06
4.25 3.06
3.95 3.66
4.25 3.66
2.75 2.46
3.05 2.46
2.75 3.06
3.05 3.06
2.75 3.66
3.05 3.66
};
\addplot [indigo68184, opacity=0.5, mark=pentagon*, mark size=2.73861278752583, mark options={solid,draw=black}, only marks]
table {%
3.43181818181818 3.11454545454545
};
\addplot [darkorange25512714, mark=*, mark size=2.73861278752583, mark options={solid,fill=darkslateblue6857130}, only marks]
table {%
6.65 2.46
};
\addplot [darkslateblue6857130, opacity=0.2, mark=*, mark size=2.73861278752583, mark options={solid}, only marks]
table {%
7.55 0.66
7.85 0.66
7.55 1.26
7.85 1.26
7.55 1.86
7.85 1.86
7.55 2.46
7.85 2.46
7.55 3.06
7.85 3.06
7.55 3.66
7.85 3.66
6.35 0.66
6.65 0.66
6.35 1.26
6.65 1.26
6.35 1.86
6.65 1.86
6.35 2.46
6.65 2.46
6.35 3.06
6.65 3.06
6.35 3.66
6.65 3.66
};
\addplot [darkslateblue6857130, opacity=0.5, mark=pentagon*, mark size=2.73861278752583, mark options={solid,draw=black}, only marks]
table {%
7.1 2.16
};
\addplot [forestgreen4416044, mark=*, mark size=2.73861278752583, mark options={solid,fill=darkslateblue48103141}, only marks]
table {%
3.05 1.26
};
\addplot [darkslateblue48103141, opacity=0.2, mark=*, mark size=2.73861278752583, mark options={solid}, only marks]
table {%
3.95 0.66
4.25 0.66
3.95 1.26
4.25 1.26
3.95 1.86
2.75 0.66
3.05 0.66
2.75 1.26
3.05 1.26
2.75 1.86
3.05 1.86
};
\addplot [darkslateblue48103141, opacity=0.5, mark=pentagon*, mark size=2.73861278752583, mark options={solid,draw=black}, only marks]
table {%
3.43181818181818 1.20545454545455
};
\addplot [crimson2143940, mark=*, mark size=2.73861278752583, mark options={solid,fill=darkcyan32144140}, only marks]
table {%
0.65 3.06
};
\addplot [darkcyan32144140, opacity=0.2, mark=*, mark size=2.73861278752583, mark options={solid}, only marks]
table {%
1.55 2.46
1.85 2.46
1.55 3.06
1.85 3.06
1.55 3.66
1.85 3.66
0.35 2.46
0.65 2.46
0.35 3.06
0.65 3.06
0.35 3.66
0.65 3.66
};
\addplot [darkcyan32144140, opacity=0.5, mark=pentagon*, mark size=2.73861278752583, mark options={solid,draw=black}, only marks]
table {%
1.1 3.06
};
\addplot [mediumpurple148103189, mark=*, mark size=2.73861278752583, mark options={solid,fill=mediumseagreen53183120}, only marks]
table {%
5.15 2.46
};
\addplot [mediumseagreen53183120, opacity=0.2, mark=*, mark size=2.73861278752583, mark options={solid}, only marks]
table {%
5.15 0.66
5.45 0.66
5.15 1.26
5.45 1.26
5.15 1.86
5.45 1.86
5.15 2.46
5.45 2.46
5.15 3.06
5.45 3.06
5.15 3.66
5.45 3.66
4.25 1.86
4.25 2.46
};
\addplot [mediumseagreen53183120, opacity=0.5, mark=pentagon*, mark size=2.73861278752583, mark options={solid,draw=black}, only marks]
table {%
5.15 2.16
};
\addplot [sienna1408675, mark=*, mark size=2.73861278752583, mark options={solid,fill=yellowgreen14421467}, only marks]
table {%
0.65 1.26
};
\addplot [yellowgreen14421467, opacity=0.2, mark=*, mark size=2.73861278752583, mark options={solid}, only marks]
table {%
1.55 0.66
1.85 0.66
1.55 1.26
1.85 1.26
1.55 1.86
1.85 1.86
0.35 0.66
0.65 0.66
0.35 1.26
0.65 1.26
0.35 1.86
0.65 1.86
};
\addplot [yellowgreen14421467, opacity=0.5, mark=pentagon*, mark size=2.73861278752583, mark options={solid,draw=black}, only marks]
table {%
1.1 1.26
};
\end{axis}

\end{tikzpicture}

%% file: figures/tx-powers-per-M.tex
\begin{tikzpicture}

\definecolor{darkgray176}{RGB}{176,176,176}
\definecolor{steelblue31119180}{RGB}{31,119,180}

\begin{axis}[
tick align=outside,
tick pos=left,
width=0.98\linewidth,
height=0.5\linewidth,
legend style={
  fill opacity=0.8,
  draw opacity=1,
  text opacity=1,
  at={(0.97,0.37)},
  anchor=south east,
},
x grid style={darkgray176},
xlabel={Number of antennas \(M\)},
xmin=-2.1, xmax=88.1,
xtick style={color=black},
y grid style={darkgray176},
ylabel={Power \si{dBm}},
ymin=10*log10(0.932814289215744*1000), ymax=10*log10(6.06777525648871*1000),
ytick style={color=black},
legend cell align={left},
y filter/.code={\pgfmathparse{10*log10(#1*1000)}} %
]

\addplot [thick, steelblue31119180] table [col sep=space, x=x, y=y] {figures/techtile/non-coherent/partial-power-tx.dat};\addlegendentry{Non-coherent system}

\addplot [dashed, thick, steelblue31119180]
table {%
2 5.70450434000166
3 4.59182094702432
4 4.18911067547895
5 3.98083544502177
6 3.77256278718074
7 3.58117848415027
8 3.39677800857025
9 3.22601725817211
10 3.0799816405635
11 2.96430033449761
12 3.04894725505044
13 2.84267261287768
14 2.71594493149203
15 2.66869090298275
16 2.56795788508226
17 2.41322034817489
18 2.40561729044
19 2.48733763152412
20 2.44957587580815
21 2.27671344290694
22 2.20060633771136
23 2.17695160667049
24 2.19472999649569
25 2.11673066612136
26 2.14542088785135
27 2.0894182257116
28 2.08557301291239
29 2.02156566820059
30 1.97382477074715
31 1.99866593955311
32 1.86617757329204
33 1.90669690185834
34 1.84045316886522
35 1.80666978697501
36 1.83494781846967
37 1.7950194887301
38 1.76652907935848
39 1.72539232219129
40 1.70606060035821
41 1.70554045071784
42 1.64970683659168
43 1.65238707193779
44 1.61157793016197
45 1.60210208268522
46 1.5938062782165
47 1.53585877611809
48 1.60173175027246
49 1.56243713172761
50 1.53611293338147
51 1.52708229539537
52 1.48441035241533
53 1.4718799662104
54 1.51237987866247
55 1.46692177894558
56 1.46776806455471
57 1.43956460163011
58 1.42172199818366
59 1.41553986849851
60 1.40883169482632
61 1.3908636943005
62 1.38062778328286
63 1.35462926645493
64 1.36253195671475
65 1.36840952720292
66 1.3397687001952
67 1.34883224952686
68 1.31271817881741
69 1.29151664499485
70 1.3129255637823
71 1.2848045383652
72 1.26186530563498
73 1.30044018592381
74 1.26676150038961
75 1.25609372359774
76 1.25677389632569
77 1.24344576317646
78 1.21347878002347
79 1.21081821407724
80 1.22208321300998
81 1.20680173939996
82 1.21079652541649
83 1.18427488333657
84 1.16753921841345
};
\addlegendentry{Coherent system}
\end{axis}

\end{tikzpicture}

%% file: figures/techtile/rx-cdf-total.tex
\begin{tikzpicture}

\definecolor{crimson2143940}{RGB}{214,39,40}
\definecolor{darkgray176}{RGB}{176,176,176}
\definecolor{darkorange25512714}{RGB}{255,127,14}
\definecolor{forestgreen4416044}{RGB}{44,160,44}
\definecolor{lightgray204}{RGB}{204,204,204}
\definecolor{mediumpurple148103189}{RGB}{148,103,189}
\definecolor{orchid227119194}{RGB}{227,119,194}
\definecolor{sienna1408675}{RGB}{140,86,75}
\definecolor{steelblue31119180}{RGB}{31,119,180}

\begin{axis}[
legend cell align={right},
legend style={
  fill opacity=0.8,
  draw opacity=1,
  text opacity=1,
  at={(0.97,0.03)},
  anchor=south east,
  draw=lightgray204
},
width=0.9\linewidth,
height=0.4\linewidth,
tick align=outside,
tick pos=left,
xlabel={Received DC energy [\si{\joule}]},
xmin=-0.791440786011763, xmax=20.6202565061858, %
xtick style={color=black},
ylabel={CDF},
ymin=-0.0497916666666667, ymax=1.045625,
ytick style={color=black},
legend columns=5,
]

\addlegendimage{steelblue31119180, only marks,mark=*}\addlegendentry{$2$}
\addlegendimage{darkorange25512714, only marks,mark=*}\addlegendentry{$8$}
\addlegendimage{forestgreen4416044, only marks,mark=*}\addlegendentry{$84$}

\addlegendimage{black}\addlegendentry{Non-coh.}
\addlegendimage{black, dashed}\addlegendentry{Coh.}

\input{figures/techtile/rx-cdf-non-coherent}
\input{figures/techtile/rx-cdf-coherent}

\end{axis}

\end{tikzpicture}

%% file: figures/techtile/non-coherent/measurements.tex
\begin{tikzpicture}

\definecolor{darkgray176}{RGB}{176,176,176}
\definecolor{darkorange25512714}{RGB}{255,127,14}
\definecolor{forestgreen4416044}{RGB}{44,160,44}
\definecolor{lightgray204}{RGB}{204,204,204}
\definecolor{steelblue31119180}{RGB}{31,119,180}

\begin{axis}[
legend cell align={left},
legend style={
  fill opacity=0.8,
  draw opacity=1,
  text opacity=1,
  at={(0.97,0.03)},
  anchor=south east,
  draw=lightgray204
},
width=0.9\linewidth,
height=0.4\linewidth,
tick align=outside,
tick pos=left,
x grid style={darkgray176},
xlabel={Received DC energy [\si{\milli\joule}]},
xmajorgrids,
xmin=-0.0706726881942274*(12 * 60 * 60/1e3), xmax=1.62721067423401*(12 * 60 * 60/1e3),
xtick style={color=black},
y grid style={darkgray176},
ylabel={CDF},
ymajorgrids,
ymin=-0.0497916666666667, ymax=1.045625,
ytick style={color=black},
x filter/.code={\pgfmathparse{#1 *(12 * 60 * 60/1e3)}} %
]
\addplot [thick, steelblue31119180]
table {%
0.00650382827978352 0
0.0108691696165292 0.00416666666666667
0.014599104830337 0.00833333333333333
0.01503174818381 0.0125
0.0152735732939663 0.0166666666666667
0.0157988441367011 0.0208333333333333
0.0181532607089972 0.025
0.0193679791755007 0.0291666666666667
0.0206105203807814 0.0333333333333333
0.0220660245204424 0.0375
0.0239825703455139 0.0416666666666667
0.024756140487884 0.0458333333333333
0.0258963362442747 0.05
0.0286537681110736 0.0541666666666667
0.0294536109138565 0.0583333333333333
0.0296090047547572 0.0625
0.0297668037513637 0.0666666666666667
0.0300263322900416 0.0708333333333333
0.0312227201164164 0.075
0.0361864247438113 0.0791666666666667
0.0364094935548675 0.0833333333333333
0.0371066534970119 0.0875
0.0375725948947329 0.0916666666666667
0.0377131097552135 0.0958333333333333
0.0377439874080011 0.1
0.0407512303065596 0.104166666666667
0.0441196775023265 0.108333333333333
0.0487614676305086 0.1125
0.0487614676305086 0.116666666666667
0.0487614676305086 0.120833333333333
0.0487614676305086 0.125
0.0537787725458297 0.129166666666667
0.0592383755000201 0.133333333333333
0.059376121173203 0.1375
0.0628394535655774 0.141666666666667
0.0661027028772224 0.145833333333333
0.0700787695117593 0.15
0.0779010808679699 0.154166666666667
0.0779421776213393 0.158333333333333
0.0785393482501397 0.1625
0.0788205607407358 0.166666666666667
0.0795616778576987 0.170833333333333
0.0829343042503182 0.175
0.0852417435739684 0.179166666666667
0.0895521300930848 0.183333333333333
0.0899301721872952 0.1875
0.0899301721872952 0.191666666666667
0.0899301721872952 0.195833333333333
0.0899301721872952 0.2
0.0899301721872952 0.204166666666667
0.0899301721872952 0.208333333333333
0.0899301721872952 0.2125
0.0933332048416195 0.216666666666667
0.0948310616146254 0.220833333333333
0.0952503007171621 0.225
0.0956838312738055 0.229166666666667
0.0983131936533548 0.233333333333333
0.098669800019396 0.2375
0.0986720945067894 0.241666666666667
0.0988466871730413 0.245833333333333
0.100154246451826 0.25
0.109273558016785 0.254166666666667
0.110439768437236 0.258333333333333
0.11356678319853 0.2625
0.113637889035606 0.266666666666667
0.120591252898265 0.270833333333333
0.125814929433319 0.275
0.129138665645653 0.279166666666667
0.132273403080265 0.283333333333333
0.132281580087982 0.2875
0.132433209135476 0.291666666666667
0.134402317317639 0.295833333333333
0.137228286302147 0.3
0.139132180875336 0.304166666666667
0.14097181780377 0.308333333333333
0.149268377762397 0.3125
0.153633424804383 0.316666666666667
0.154765286396589 0.320833333333333
0.156015365507901 0.325
0.156411602394579 0.329166666666667
0.161826401966464 0.333333333333333
0.164418944041219 0.3375
0.164859140603291 0.341666666666667
0.167469415615305 0.345833333333333
0.170403256725777 0.35
0.174646629148397 0.354166666666667
0.176702129675112 0.358333333333333
0.185969455892456 0.3625
0.187646925708834 0.366666666666667
0.188245711615389 0.370833333333333
0.190520660944911 0.375
0.19434025840739 0.379166666666667
0.198675728969821 0.383333333333333
0.20328093359983 0.3875
0.205463490864341 0.391666666666667
0.205498807165028 0.395833333333333
0.211356503780492 0.4
0.214925559041685 0.404166666666667
0.223224979462402 0.408333333333333
0.231530306530835 0.4125
0.231581004352672 0.416666666666667
0.233826952599036 0.420833333333333
0.234980125735553 0.425
0.235154354408066 0.429166666666667
0.236283214401868 0.433333333333333
0.238188695267591 0.4375
0.247296071961809 0.441666666666667
0.247296071961809 0.445833333333333
0.247296071961809 0.45
0.247296071961809 0.454166666666667
0.247296071961809 0.458333333333333
0.247296071961809 0.4625
0.247393414774256 0.466666666666667
0.247393414774256 0.470833333333333
0.247393414774256 0.475
0.247393414774256 0.479166666666667
0.247393414774256 0.483333333333333
0.247393414774256 0.4875
0.247393414774256 0.491666666666667
0.247393414774256 0.495833333333333
0.247393414774256 0.5
0.247393414774256 0.504166666666667
0.247393414774256 0.508333333333333
0.247393414774256 0.5125
0.247393414774256 0.516666666666667
0.247393414774256 0.520833333333333
0.251806249921934 0.525
0.256568513398315 0.529166666666667
0.257513672416583 0.533333333333333
0.27096778821526 0.5375
0.272748893906942 0.541666666666667
0.272827716518875 0.545833333333333
0.279648242954253 0.55
0.279648242954253 0.554166666666667
0.279648242954253 0.558333333333333
0.279648242954253 0.5625
0.279648242954253 0.566666666666667
0.279648242954253 0.570833333333333
0.281156003451845 0.575
0.285903977496005 0.579166666666667
0.290777899842337 0.583333333333333
0.294205443836402 0.5875
0.300689855365767 0.591666666666667
0.301100725772176 0.595833333333333
0.301100725772176 0.6
0.301100725772176 0.604166666666667
0.301100725772176 0.608333333333333
0.301100725772176 0.6125
0.301100725772176 0.616666666666667
0.301100725772176 0.620833333333333
0.301100725772176 0.625
0.301100725772176 0.629166666666667
0.301100725772176 0.633333333333333
0.301100725772176 0.6375
0.301100725772176 0.641666666666667
0.301100725772176 0.645833333333333
0.301100725772176 0.65
0.301100725772176 0.654166666666667
0.301100725772176 0.658333333333333
0.301100725772176 0.6625
0.301100725772176 0.666666666666667
0.301100725772176 0.670833333333333
0.301100725772176 0.675
0.301100725772176 0.679166666666667
0.301100725772176 0.683333333333333
0.301100725772176 0.6875
0.301100725772176 0.691666666666667
0.301100725772176 0.695833333333333
0.301100725772176 0.7
0.301100725772176 0.704166666666667
0.301100725772176 0.708333333333333
0.301100725772176 0.7125
0.301100725772176 0.716666666666667
0.301100725772176 0.720833333333333
0.301100725772176 0.725
0.301100725772176 0.729166666666667
0.301100725772176 0.733333333333333
0.305527557587801 0.7375
0.306251171009299 0.741666666666667
0.30693067533626 0.745833333333333
0.315600416648735 0.75
0.318884692122003 0.754166666666667
0.318884692122003 0.758333333333333
0.318884692122003 0.7625
0.318884692122003 0.766666666666667
0.318884692122003 0.770833333333333
0.318884692122003 0.775
0.318884692122003 0.779166666666667
0.318884692122003 0.783333333333333
0.318884692122003 0.7875
0.318884692122003 0.791666666666667
0.318884692122003 0.795833333333333
0.318884692122003 0.8
0.318884692122003 0.804166666666667
0.318884692122003 0.808333333333333
0.318884692122003 0.8125
0.318884692122003 0.816666666666667
0.318884692122003 0.820833333333333
0.318884692122003 0.825
0.318884692122003 0.829166666666667
0.318884692122003 0.833333333333333
0.318884692122003 0.8375
0.318884692122003 0.841666666666667
0.318884692122003 0.845833333333333
0.318884692122003 0.85
0.322192159918101 0.854166666666667
0.32361406165257 0.858333333333333
0.358582663228837 0.8625
0.358582663228837 0.866666666666667
0.358582663228837 0.870833333333333
0.358582663228837 0.875
0.40097549846777 0.879166666666667
0.415682054388698 0.883333333333333
0.415682054388698 0.8875
0.415682054388698 0.891666666666667
0.415682054388698 0.895833333333333
0.415682054388698 0.9
0.427988505985736 0.904166666666667
0.443350252554665 0.908333333333333
0.47701608659899 0.9125
0.477502756510862 0.916666666666667
0.477502756510862 0.920833333333333
0.477502756510862 0.925
0.477502756510862 0.929166666666667
0.477502756510862 0.933333333333333
0.481598778086655 0.9375
0.509440120392221 0.941666666666667
0.51835137334443 0.945833333333333
0.57094103412226 0.95
0.590370702067295 0.954166666666667
0.665669263144018 0.958333333333333
0.674626860252193 0.9625
0.743124257031384 0.966666666666667
0.743124257031384 0.970833333333333
0.743124257031384 0.975
0.743124257031384 0.979166666666667
1.55003415776 0.983333333333333
1.55003415776 0.9875
1.55003415776 0.991666666666667
1.55003415776 0.995833333333333
};
\addlegendentry{2 (\SI{14.76}{dBm})}
\addplot [thick, darkorange25512714]
table {%
0.0148918082862586 0
0.0203666845439006 0.00416666666666667
0.0242629511384633 0.00833333333333333
0.0267525786269767 0.0125
0.0285353805055491 0.0166666666666667
0.0316398806747034 0.0208333333333333
0.0319276969340377 0.025
0.0328638342667835 0.0291666666666667
0.0342325046252061 0.0333333333333333
0.0345547707988727 0.0375
0.0358501329448646 0.0416666666666667
0.035924204307507 0.0458333333333333
0.0360565853727758 0.05
0.0373609220127163 0.0541666666666667
0.0381330825329313 0.0583333333333333
0.0404695063019256 0.0625
0.0408663664143623 0.0666666666666667
0.0409970596291626 0.0708333333333333
0.0416031385389341 0.075
0.0417154627897619 0.0791666666666667
0.0425496629534186 0.0833333333333333
0.0445465198540914 0.0875
0.0446406675472205 0.0916666666666667
0.0454650783054529 0.0958333333333333
0.0467251768500469 0.1
0.0470518848884239 0.104166666666667
0.0486242230824742 0.108333333333333
0.0493122098689081 0.1125
0.05123343972324 0.116666666666667
0.0512637007241503 0.120833333333333
0.0522747496217142 0.125
0.0525449879271534 0.129166666666667
0.0526526337426447 0.133333333333333
0.0526548854902707 0.1375
0.0532745611761929 0.141666666666667
0.0541308009818164 0.145833333333333
0.054733078083736 0.15
0.054792322867115 0.154166666666667
0.0555252056788854 0.158333333333333
0.0570319424609071 0.1625
0.0592842354019233 0.166666666666667
0.0595682034536811 0.170833333333333
0.0596519665584528 0.175
0.0598305476402887 0.179166666666667
0.0610873679238651 0.183333333333333
0.0611982248293304 0.1875
0.0622663861694545 0.191666666666667
0.0625323760514497 0.195833333333333
0.0648766537466114 0.2
0.0660914460148446 0.204166666666667
0.0678461483143099 0.208333333333333
0.0680645082157093 0.2125
0.0687541056419016 0.216666666666667
0.0689395582621335 0.220833333333333
0.0689395582621335 0.225
0.0689395582621335 0.229166666666667
0.0689395582621335 0.233333333333333
0.0689395582621335 0.2375
0.0690661140833613 0.241666666666667
0.0696048854774234 0.245833333333333
0.0702375880614127 0.25
0.0709181029339429 0.254166666666667
0.0722695132291568 0.258333333333333
0.0737155813805861 0.2625
0.0753784644341788 0.266666666666667
0.0753784644341788 0.270833333333333
0.0753784644341788 0.275
0.0753784644341788 0.279166666666667
0.0753784644341788 0.283333333333333
0.0753784644341788 0.2875
0.0753784644341788 0.291666666666667
0.0753784644341788 0.295833333333333
0.0753784644341788 0.3
0.0753784644341788 0.304166666666667
0.0753784644341788 0.308333333333333
0.0753784644341788 0.3125
0.0753784644341788 0.316666666666667
0.0753784644341788 0.320833333333333
0.0753784644341788 0.325
0.0777592229803181 0.329166666666667
0.0792962998401543 0.333333333333333
0.0792962998401543 0.3375
0.0792962998401543 0.341666666666667
0.0792962998401543 0.345833333333333
0.0792962998401543 0.35
0.0792962998401543 0.354166666666667
0.0792962998401543 0.358333333333333
0.0792962998401543 0.3625
0.0792962998401543 0.366666666666667
0.0792962998401543 0.370833333333333
0.0792962998401543 0.375
0.0792962998401543 0.379166666666667
0.0792962998401543 0.383333333333333
0.0792962998401543 0.3875
0.0792962998401543 0.391666666666667
0.0804372399927132 0.395833333333333
0.0807858356911798 0.4
0.0813474681856311 0.404166666666667
0.0815463726360813 0.408333333333333
0.0819818046896722 0.4125
0.0827731263688882 0.416666666666667
0.0880632256211481 0.420833333333333
0.0880632256211481 0.425
0.0880632256211481 0.429166666666667
0.0880632256211481 0.433333333333333
0.0880632256211481 0.4375
0.0880632256211481 0.441666666666667
0.0880632256211481 0.445833333333333
0.0880632256211481 0.45
0.0880632256211481 0.454166666666667
0.0880632256211481 0.458333333333333
0.0880632256211481 0.4625
0.0880632256211481 0.466666666666667
0.0880632256211481 0.470833333333333
0.0880632256211481 0.475
0.0880632256211481 0.479166666666667
0.0880632256211481 0.483333333333333
0.0880632256211481 0.4875
0.0880632256211481 0.491666666666667
0.0880632256211481 0.495833333333333
0.0880632256211481 0.5
0.0880632256211481 0.504166666666667
0.0880632256211481 0.508333333333333
0.0880632256211481 0.5125
0.0880632256211481 0.516666666666667
0.0880632256211481 0.520833333333333
0.0881100542533445 0.525
0.0888607715828201 0.529166666666667
0.0913751500120515 0.533333333333333
0.0931014484861423 0.5375
0.0971147386916781 0.541666666666667
0.0985233661409529 0.545833333333333
0.0985233661409529 0.55
0.0985406309473006 0.554166666666667
0.0985406309473006 0.558333333333333
0.0985406309473006 0.5625
0.0985406309473006 0.566666666666667
0.0985406309473006 0.570833333333333
0.0985406309473006 0.575
0.0985406309473006 0.579166666666667
0.099854899993194 0.583333333333333
0.104332349574996 0.5875
0.108856245002289 0.591666666666667
0.110786273672579 0.595833333333333
0.114940761643252 0.6
0.122443607151355 0.604166666666667
0.123460969574344 0.608333333333333
0.124659558584502 0.6125
0.124659558584502 0.616666666666667
0.124659558584502 0.620833333333333
0.124659558584502 0.625
0.124659558584502 0.629166666666667
0.124659558584502 0.633333333333333
0.124659558584502 0.6375
0.124659558584502 0.641666666666667
0.124659558584502 0.645833333333333
0.124659558584502 0.65
0.131591924986307 0.654166666666667
0.13395292393563 0.658333333333333
0.136476479802876 0.6625
0.144960716079549 0.666666666666667
0.144960716079549 0.670833333333333
0.144960716079549 0.675
0.144960716079549 0.679166666666667
0.144960716079549 0.683333333333333
0.144960716079549 0.6875
0.144960716079549 0.691666666666667
0.144960716079549 0.695833333333333
0.144960716079549 0.7
0.144960716079549 0.704166666666667
0.144960716079549 0.708333333333333
0.144960716079549 0.7125
0.144960716079549 0.716666666666667
0.144960716079549 0.720833333333333
0.144960716079549 0.725
0.144960716079549 0.729166666666667
0.144960716079549 0.733333333333333
0.144960716079549 0.7375
0.144960716079549 0.741666666666667
0.144960716079549 0.745833333333333
0.144960716079549 0.75
0.144960716079549 0.754166666666667
0.144960716079549 0.758333333333333
0.144960716079549 0.7625
0.144960716079549 0.766666666666667
0.14714191498728 0.770833333333333
0.148207924611329 0.775
0.14940784582902 0.779166666666667
0.157202318408367 0.783333333333333
0.163464023793332 0.7875
0.170269726480582 0.791666666666667
0.171701354600395 0.795833333333333
0.172148297892106 0.8
0.173103854335498 0.804166666666667
0.176402827986036 0.808333333333333
0.178646153078288 0.8125
0.183704206717948 0.816666666666667
0.203577833083333 0.820833333333333
0.206219300979108 0.825
0.21147315008521 0.829166666666667
0.218905362518714 0.833333333333333
0.2236201608552 0.8375
0.257669416017272 0.841666666666667
0.257669416017272 0.845833333333333
0.257669416017272 0.85
0.257669416017272 0.854166666666667
0.257669416017272 0.858333333333333
0.257669416017272 0.8625
0.257669416017272 0.866666666666667
0.257669416017272 0.870833333333333
0.257669416017272 0.875
0.257669416017272 0.879166666666667
0.257669416017272 0.883333333333333
0.257669416017272 0.8875
0.257669416017272 0.891666666666667
0.257669416017272 0.895833333333333
0.257669416017272 0.9
0.259581372843361 0.904166666666667
0.269770704662977 0.908333333333333
0.282199997900177 0.9125
0.283048409377783 0.916666666666667
0.283591046907787 0.920833333333333
0.290624720616244 0.925
0.292277659744047 0.929166666666667
0.323040826792233 0.933333333333333
0.348297440115717 0.9375
0.35994565584225 0.941666666666667
0.35994565584225 0.945833333333333
0.35994565584225 0.95
0.426096440116598 0.954166666666667
0.449934382961483 0.958333333333333
0.481029144322862 0.9625
0.492308532996549 0.966666666666667
0.541955993444083 0.970833333333333
0.546670084356611 0.975
0.623856054420397 0.979166666666667
0.755107008882092 0.983333333333333
1.0403874723316 0.9875
1.05547452560807 0.991666666666667
1.42614232697156 0.995833333333333
};
\addlegendentry{8 (\SI{13.10}{dBm})}
\addplot [thick, forestgreen4416044]
table {%
0.00995709485085202 0
0.014197425604924 0.00416666666666667
0.014197425604924 0.00833333333333333
0.014197425604924 0.0125
0.0176640082706568 0.0166666666666667
0.0186608721339532 0.0208333333333333
0.0217627013274819 0.025
0.0244497401661564 0.0291666666666667
0.0257089972688464 0.0333333333333333
0.0258644465472825 0.0375
0.0281097934463112 0.0416666666666667
0.0303146221931731 0.0458333333333333
0.0312746387621612 0.05
0.0332233636594524 0.0541666666666667
0.0332233636594524 0.0583333333333333
0.0332233636594524 0.0625
0.0332258762627598 0.0666666666666667
0.0356950535304184 0.0708333333333333
0.0373415881397065 0.075
0.0380394242281161 0.0791666666666667
0.0389871868165817 0.0833333333333333
0.0393234646644118 0.0875
0.0405393031801677 0.0916666666666667
0.0417887685396648 0.0958333333333333
0.0421256399683933 0.1
0.0447452966985817 0.104166666666667
0.0452639761217606 0.108333333333333
0.0463802286336789 0.1125
0.0466878267676656 0.116666666666667
0.0468294768520166 0.120833333333333
0.0479831094130946 0.125
0.0480174373166974 0.129166666666667
0.0484131058751687 0.133333333333333
0.0488245302520965 0.1375
0.0494042600349883 0.141666666666667
0.0513586605834233 0.145833333333333
0.0513586605834233 0.15
0.0513586605834233 0.154166666666667
0.0513586605834233 0.158333333333333
0.054938078470603 0.1625
0.0551263859766487 0.166666666666667
0.0552771878619107 0.170833333333333
0.0564114476584649 0.175
0.0575102761064608 0.179166666666667
0.0597157598708842 0.183333333333333
0.0603755422367768 0.1875
0.0609328475245481 0.191666666666667
0.0616814710424324 0.195833333333333
0.0631661408644394 0.2
0.0635421623620483 0.204166666666667
0.0643166434456422 0.208333333333333
0.0656302694627357 0.2125
0.0658996254224934 0.216666666666667
0.0661652118990512 0.220833333333333
0.0663743461451777 0.225
0.067499646214298 0.229166666666667
0.0679330627866144 0.233333333333333
0.0692287190582571 0.2375
0.0694127424493592 0.241666666666667
0.0710975994758694 0.245833333333333
0.0715231144010472 0.25
0.0716223316591743 0.254166666666667
0.0719547582546321 0.258333333333333
0.0722700183185589 0.2625
0.072609932848803 0.266666666666667
0.0736789342676797 0.270833333333333
0.0743408802405674 0.275
0.0769911003769014 0.279166666666667
0.0773915372403888 0.283333333333333
0.0779786684070277 0.2875
0.0802328945970801 0.291666666666667
0.0829636692412821 0.295833333333333
0.084567663933198 0.3
0.0849623041119902 0.304166666666667
0.0853222206996743 0.308333333333333
0.0877958433873299 0.3125
0.0905755367223801 0.316666666666667
0.0915793562954524 0.320833333333333
0.0915793562954524 0.325
0.0915793562954524 0.329166666666667
0.0915793562954524 0.333333333333333
0.0915793562954524 0.3375
0.0915793562954524 0.341666666666667
0.0915793562954524 0.345833333333333
0.0915793562954524 0.35
0.0915793562954524 0.354166666666667
0.0915793562954524 0.358333333333333
0.0915793562954524 0.3625
0.0915793562954524 0.366666666666667
0.0915793562954524 0.370833333333333
0.0915793562954524 0.375
0.0915793562954524 0.379166666666667
0.0915793562954524 0.383333333333333
0.0915793562954524 0.3875
0.0920013826863848 0.391666666666667
0.0956939209905193 0.395833333333333
0.0956939209905193 0.4
0.0956939209905193 0.404166666666667
0.0956939209905193 0.408333333333333
0.0956939209905193 0.4125
0.0956939209905193 0.416666666666667
0.0956939209905193 0.420833333333333
0.0956939209905193 0.425
0.0956939209905193 0.429166666666667
0.0956939209905193 0.433333333333333
0.0956939209905193 0.4375
0.0956939209905193 0.441666666666667
0.0967529651778841 0.445833333333333
0.0967529651778841 0.45
0.0967529651778841 0.454166666666667
0.0967529651778841 0.458333333333333
0.0967529651778841 0.4625
0.0967529651778841 0.466666666666667
0.0974456403772083 0.470833333333333
0.0974456403772083 0.475
0.100143516402115 0.479166666666667
0.101404869449851 0.483333333333333
0.106419879265159 0.4875
0.106419879265159 0.491666666666667
0.106419879265159 0.495833333333333
0.106419879265159 0.5
0.106419879265159 0.504166666666667
0.106419879265159 0.508333333333333
0.106419879265159 0.5125
0.107326663415149 0.516666666666667
0.109360759205956 0.520833333333333
0.109693269978789 0.525
0.110457455182802 0.529166666666667
0.111423061147173 0.533333333333333
0.113932469474395 0.5375
0.114898715377768 0.541666666666667
0.115522817578796 0.545833333333333
0.115828863910093 0.55
0.1171306923614 0.554166666666667
0.117917495716702 0.558333333333333
0.121337651174259 0.5625
0.122085671983741 0.566666666666667
0.123226162441349 0.570833333333333
0.123645501936175 0.575
0.123776332059302 0.579166666666667
0.125369951711967 0.583333333333333
0.130214943216097 0.5875
0.130761683575908 0.591666666666667
0.130761683575908 0.595833333333333
0.130761683575908 0.6
0.130761683575908 0.604166666666667
0.132724749880653 0.608333333333333
0.136155719897247 0.6125
0.136250610141441 0.616666666666667
0.136975618610367 0.620833333333333
0.138072432087503 0.625
0.13956630148875 0.629166666666667
0.145169987608811 0.633333333333333
0.145285926252748 0.6375
0.145285926252748 0.641666666666667
0.145285926252748 0.645833333333333
0.145285926252748 0.65
0.145285926252748 0.654166666666667
0.145285926252748 0.658333333333333
0.145285926252748 0.6625
0.145285926252748 0.666666666666667
0.145285926252748 0.670833333333333
0.145285926252748 0.675
0.145285926252748 0.679166666666667
0.145285926252748 0.683333333333333
0.145285926252748 0.6875
0.145285926252748 0.691666666666667
0.145285926252748 0.695833333333333
0.145285926252748 0.7
0.147248973550382 0.704166666666667
0.149371547632003 0.708333333333333
0.149371547632003 0.7125
0.149371547632003 0.716666666666667
0.149371547632003 0.720833333333333
0.149371547632003 0.725
0.149371547632003 0.729166666666667
0.149371547632003 0.733333333333333
0.149371547632003 0.7375
0.149371547632003 0.741666666666667
0.149371547632003 0.745833333333333
0.149371547632003 0.75
0.149371547632003 0.754166666666667
0.149371547632003 0.758333333333333
0.149371547632003 0.7625
0.149371547632003 0.766666666666667
0.149371547632003 0.770833333333333
0.149371547632003 0.775
0.151753065122434 0.779166666666667
0.152494700784818 0.783333333333333
0.154646502012207 0.7875
0.164535382877894 0.791666666666667
0.181775859236404 0.795833333333333
0.181891936612128 0.8
0.182263981515857 0.804166666666667
0.182377491431786 0.808333333333333
0.18603429700275 0.8125
0.187350978157822 0.816666666666667
0.189218721217636 0.820833333333333
0.192994256188916 0.825
0.193085260692725 0.829166666666667
0.193085260692725 0.833333333333333
0.194272532582812 0.8375
0.194272532582812 0.841666666666667
0.194272532582812 0.845833333333333
0.194272532582812 0.85
0.195750796901813 0.854166666666667
0.202873006788727 0.858333333333333
0.235366140219109 0.8625
0.238064613411754 0.866666666666667
0.240320923863285 0.870833333333333
0.257303908161742 0.875
0.269906443409601 0.879166666666667
0.269906443409601 0.883333333333333
0.270334880503661 0.8875
0.270334880503661 0.891666666666667
0.270334880503661 0.895833333333333
0.270334880503661 0.9
0.270334880503661 0.904166666666667
0.270334880503661 0.908333333333333
0.270334880503661 0.9125
0.270334880503661 0.916666666666667
0.270334880503661 0.920833333333333
0.270334880503661 0.925
0.270334880503661 0.929166666666667
0.270334880503661 0.933333333333333
0.270334880503661 0.9375
0.270334880503661 0.941666666666667
0.270334880503661 0.945833333333333
0.270334880503661 0.95
0.270334880503661 0.954166666666667
0.270334880503661 0.958333333333333
0.270334880503661 0.9625
0.270334880503661 0.966666666666667
0.270334880503661 0.970833333333333
0.280325491129644 0.975
0.288101929375097 0.979166666666667
0.301904158732755 0.983333333333333
0.354583879995401 0.9875
0.468549648100921 0.991666666666667
0.662043082221159 0.995833333333333
};
\addlegendentry{84 (\SI{12.94}{dBm})}
\end{axis}

\end{tikzpicture}

%% file: figures/techtile/coherent/tx-powers-per-N.tex
\begin{tikzpicture}

\definecolor{darkgray176}{RGB}{176,176,176}

\begin{axis}[
colorbar,
colorbar style={title=$\|\vect{w}_n\|^2$,ylabel={dBm}},
colormap/viridis,
point meta max=63.2222061558011,
point meta min=-20,
tick align=outside,
tick pos=left,
x grid style={darkgray176},
xlabel={Time (hours)},
xmin=0, xmax=168,
xtick style={color=black},
xtick={0,42,84,126,168},
xticklabels={0,3,6,9,12},
y grid style={darkgray176},
ylabel={Number of antennas \(M\)},
ymin=0, ymax=14,
ytick style={color=black},
ytick={0.5,1.5,2.5,3.5,4.5,5.5,6.5,7.5,8.5,9.5,10.5,11.5,12.5,13.5},
yticklabels={2,3,4,6,7,8,12,14,21,24,28,42,56,84},
width=0.7\linewidth,
height=0.6\linewidth,
]
\addplot graphics [includegraphics cmd=\pgfimage,xmin=0, xmax=168, ymin=0, ymax=14] {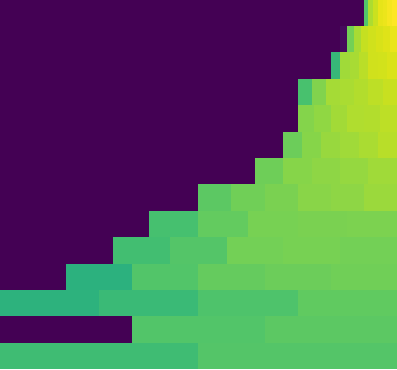};
\end{axis}

\end{tikzpicture}

%% file: sections/04_Measurements.tex
\subsection{Experimental Evaluation}\label{sec:eval-exp}
The system setup, as introduced in~\cref{sec:eval-scenario} and illustrated in~\cref{fig:techtile-setup}, is emulated in the Techtile testbed~\cite{CallebautTechtile} using all the \num{84} antennas on the ceiling. The experiment is conducted by using the results of the optimization problem in the non-coherent scenario, i.e., the power allocation per antenna. Measurements are performed at \SI{917}{\mega\hertz} by collecting the average \gls{rf} power at the \gls{esl} locations.
The measured \gls{cdf} of the received \gls{dc} energy is plotted in~\cref{fig:cdf-dc-rx-meas}, where the \gls{dc} energy is computed as $E_k = \alpha p_\text{RF}{}_{k} N T$. The minimum received power \(\beta\) is omitted, as the testbed has a lower maximum power (\SI{13.4}{dBm}) than required. Due to this limitation, the power allocations yielded by the optimization are scaled, i.e., \SI{36}{dBm} becomes \SI{13.4}{dBm}. The relative allocated transmit powers are hence kept the same. 
This yields thus the expected \gls{cdf}, following the numerical evaluation, where the same trends can be observed, i.e., reduction of required transmit power, variance, and maximum received energy with increased number of available antennas.

%% file: sections/05_conclusions.tex
\glsresetall
\section{Conclusions and Future Outlook}\label{sec:concl}
We have defined the problem of wirelessly charging \glspl{er} within a multi-antenna system, exploring both unsynchronized (non-coherent) and synchronized (coherent) configurations. Unlike previous approaches that aimed to maximize the total transferred energy, our focus was on minimizing the overall system transmit power while ensuring the required \gls{dc} energy supply.
The analysis reveals that increasing the number of antennas and transitioning from an unsynchronized to a synchronized system significantly improve system performance. This enhancement ensures that the received energy aligns more closely with the system's requirements, eliminating overshoots and reducing overall energy consumption. Additionally, we evaluated the non-coherent system using a testbed equipped with 84 antennas. The results obtained from this testbed corroborate the trends and observations identified in our numerical evaluations. Notable, providing a coherent system requires periodic calibration and additional signalling overhead in combination with higher hardware restrictions~\cite{CallebautTechtile}.

In future research, we aim to extend the minimization of total transmit power, to the total energy consumption of the system. Thereby also taking into account the energy related to the activation of \glspl{et}. Additionally, investigating the impact of deviations from synchronization assumptions, including errors in synchronization and channel estimation, will be crucial. 
Integrating antenna patterns into the channel model will offer a more realistic representation of the system. We also intend to expand the scope of measurements for the coherent system and account for energy loss resulting from channel estimation. Lastly, considering linear power amplification without minimum power constraints may offer insights, as the current observations show very low power amplification values; including minimum constraints might encourage the use of more time slots.